\documentclass[11pt]{article}

\usepackage[final]{acl}

\usepackage{helvet}
\usepackage{times}
\usepackage{latexsym}
\usepackage{booktabs} 
\usepackage{multirow}
\usepackage{multicol}
\usepackage{amsmath}
\usepackage[table]{xcolor}
\usepackage{tabularx} 
\usepackage{array}    
\usepackage[frozencache,cachedir=minted-main]{minted}
\usepackage{algorithm}
\usepackage{algorithmic}
\usepackage{float}
\usepackage{arydshln}  
\usepackage{bbding}
\usepackage{subcaption}
\usepackage[table]{xcolor}
\usepackage{colortbl}      
\usepackage{array}         
\usepackage[T1]{fontenc}

\usepackage[utf8]{inputenc}

\usepackage{microtype}

\usepackage{inconsolata}

\usepackage{graphicx}

\title{\textsc{Trace}: Evaluating Execution Efficiency of LLM-Based Code Translation}

\author{
Zhihao Gong$^1$,
Zeyu Sun$^3$,
Dong Huang$^4$,
Qingyuan Liang$^1$,
Jie M. Zhang$^5$,
Dan Hao$^{1,2}$\thanks{Corresponding author.} \\
$^1$Key Lab of HCST (PKU), MOE; SCS, Peking University, China \\
$^2$School of Electronic and Computer Engineering, PKU Shenzhen Graduate School, China \\
$^3$Institute of Software, Chinese Academy of Sciences, Beijing, China \\
$^4$National University of Singapore, Singapore \\
$^5$King's College London, London, United Kingdom \\
\{zhihaogong, liangqy\}@stu.pku.edu.cn, haodan@pku.edu.cn \\
zeyu.zys@gmail.com, dong.huang@nus.edu.sg, jie.zhang@kcl.ac.uk
}

\usepackage{mystyle}
\begin{document}
\maketitle
\begin{abstract}

    
While Large Language Models (LLMs) have substantially improved the functional correctness of code translation, the critical dimension of execution efficiency remains overlooked. We present \textbf{\textsc{Trace}}, the first benchmark to explicitly assess efficiency in LLM-translated code. \textsc{Trace} includes 1,000 efficiency-critical tasks across C++, Java, and Python, each augmented with stress tests that reveal efficiency disparities often overlooked by small-scale tests. Using \textsc{Trace}, we conduct an extensive evaluation of 28 representative LLMs and highlight several key insights: 1) Correctness and efficiency are often misaligned: the correctness leader Claude-Sonnet-4-Think achieves only moderate time efficiency, outperformed by smaller open-source LLMs such as Qwen2.5-Coder-14B-Instruct. 2) Inefficiency is both prevalent and patterned: 23.5\% of correct translations suffer from notable inefficiency, mainly arising from algorithm implementation discrepancy (11.9\%), language construct mismatch (66.4\%), and resource management inefficiency (21.7\%).
3) Inference-time prompt strategies bring only modest improvements, indicating that simple prompting alone is insufficient to improve translation efficiency. Together, our results establish execution efficiency as an essential dimension of code translation and position \textsc{Trace} as a principled foundation for efficiency-oriented evaluation. Our code and data are available at: \url{https://github.com/Albert-Gong/TRACE}.

\end{abstract}

\section{Introduction}

Code translation migrates programs across programming languages while preserving functionality. It supports critical tasks such as modernizing legacy systems, extending library interoperability, and consolidating disparate codebases into standardized stacks~\cite{pan2024lost}. As software grows in complexity, the demand for automatic code translation has become more pressing.

The remarkable proficiency of LLMs in multilingual code understanding and generation~\cite{zhao2023survey, zheng2023survey, zheng2023codegeex, jiang2024survey} has driven a paradigm shift in code translation. Prior work explores techniques such as back translation~\cite{ahmad2022summarize}, fine-tuning~\cite{zhu2024semi}, chain-of-thought~\cite{macedo2024intertrans, nitin2024spectra}, and self-repair~\cite{yang2024exploring} to further enhance translation. Collectively, these efforts significantly improve the syntactic and functional fidelity of LLM-based code translation.

\begin{figure}[t]
  \centering
  \begin{minipage}[t]{0.49\linewidth} 
    \vspace{0pt}
    \begin{bluemotivationbox}[title={\strut C++ Source Code}]
\begin{minted}{cpp}
int gcd(int a, int b) {
    if (a == 0) return b;
    if (b == 0) return a;
    int k;
    for (k = 0; ((a | b) & 1) == 0; ++k) {
        a >>= 1; b >>= 1;
    }
    while ((a & 1) == 0)
        a >>= 1;
    do {
        while ((b & 1) == 0)
            b >>= 1;
        if (a > b) 
            swap(a, b);
        b = b - a;
    } while (b != 0);
    return a << k;
}
\end{minted}
    \end{bluemotivationbox}

    \begin{bluemotivationbox}[title={\strut Execution Profile}]
\begin{minted}{markdown}
### Small Tests
- Test Input: `37, 93`
- Execution Time: **0.02 s vs. 0.02 s**

### Stress Tests
- Test Input: `2147483647, 2147483629`
- Execution Time: **11.20 s vs. 0.02 s** 
\end{minted}
    \end{bluemotivationbox}
  \end{minipage}%
  \hfill
  \begin{minipage}[t]{0.49\linewidth} 
    \vspace{0pt}
    \begin{bluemotivationbox}[title={\strut Python Translation}]
\begin{minted}{python}
def gcd(a, b):
    if a == 0: return b
    if b == 0: return a
    k = 0
    while (a | b) & 1 == 0:
        a >>= 1
        b >>= 1
        k += 1
    while a & 1 == 0:
        a >>= 1
    ''' CodeLlama-34B-Instruct-hf
    Inefficiency: 
        shift moved outside the loop.
    '''
-   while b & 1 == 0:
-       b >>= 1
-   while b != 0:
-       if a > b:
-           a, b = b, a
-       b = b - a
    ''' GPT-4o-Mini
    '''
+   while b != 0:
+       while b & 1 == 0:
+           b >>= 1
+       if a > b:
+           a, b = b, a
+       b = b - a
    return a << k
\end{minted}
    \end{bluemotivationbox}
  \end{minipage}

  \caption{A motivating example illustrating the critical discrepancy between functional correctness and execution efficiency in LLM-based code translation.}
  \label{figure:motivation_example}
\end{figure}

Despite advances in correctness, an equally critical dimension remains understudied: \textit{execution efficiency}~\cite{niu2024evaluating, vartziotis2024learn}. Execution efficiency determines how well a translated program performs, directly influencing responsiveness, scalability, and operational cost. Crucially, even functionally correct code translations must be efficient in practice; otherwise, excessive runtime overhead can make them impractical to use.

Figure~\ref{figure:motivation_example} illustrates an example of Stein's GCD algorithm~\cite{steins_gcd}, a canonical problem from the widely used \textsc{Transcoder-Test} benchmark~\cite{lachaux2020unsupervised}. When translating the C++ implementation into Python, solutions from CodeLlama-34B-Instruct-hf and GPT-4o-Mini appear indistinguishable under small-scale tests, both passing with nearly identical runtime (0.02\,s). However, on a stress-test input ($a=2147483647$, $b=2147483629$) from our generated stress tests, their performance diverges sharply: CodeLlama's translation runs over 500$\times$ slower (11.20\,s vs.\ 0.02\,s). This failure stems from an algorithmic misinterpretation, where CodeLlama moves the bitwise-shift outside the loop, discarding the per-iteration optimization central to Stein’s algorithm and thereby nullifying its efficiency.

This example underscores a key challenge in evaluating LLM-based code translation: \textit{functional correctness does not guarantee efficiency equivalence}. As observed in our study, even when the source code embodies a highly optimized implementation, and the translation faithfully preserves functional correctness, LLMs may still introduce algorithmic degradations or adopt suboptimal idioms, incurring substantial runtime overheads. However, evaluation protocols such as the small-scale tests in \textsc{Transcoder-Test} mainly focus on correctness and fail to expose such inefficiencies. This gap calls for treating execution efficiency as an explicit evaluation dimension.





To address the challenge, we present \textbf{\textsc{Trace}} (\textbf{TRA}nslated \textbf{C}ode \textbf{E}fficiency), the first benchmark designed to evaluate execution efficiency in LLM-translated code. We focus on translation tasks across C++, Java, and Python, which consistently rank among the three most popular programming languages~\cite{pypl,tiobe}. 
\textsc{Trace} studies efficiency in method-level code translation. It uses the well-established \textsc{Transcoder-Test} task setting as a controlled basis for evaluation. We target this granularity as it represents a fundamental translation unit and underpins broader workflows, including class- and repository-level translation~\cite{xue2025classeval,ibrahimzada2025alphatrans}.

\textsc{Trace} is built through a two-stage process. First, we iteratively generate stress tests to uncover latent efficiency issues beyond the reach of small-scale tests. Second, we apply efficiency-oriented filtering to retain only tasks that meaningfully expose runtime differences. The final benchmark comprises 1,000 efficiency-critical tasks, each accompanied by 10 correctness tests, 10 stress tests, and an average of 23 efficiency reference translations to support nuanced evaluation.



With \textsc{Trace}, we evaluate 28 representative LLMs and derive several key insights. 
First, correctness and efficiency are often misaligned. For example, the correctness leader Claude-Sonnet-4-Think achieves a 95.5\% pass rate but only moderate time efficiency, outperformed by smaller open-source LLMs such as Qwen2.5-Coder-14B-Instruct. Moreover, reasoning-enhanced or larger-scale LLMs do not reliably improve efficiency. We also observe directional asymmetry: while Java$\rightarrow$Python and Python$\rightarrow$Java achieve comparable correctness, models are markedly more efficient in the former.
Second, inefficiency is both widespread and patterned. Notably, 23.5\% of functionally correct translations are over $2\times$ slower or more memory-intensive than the most efficient counterparts. A manual study of 327 inefficient cases further reveals three major categories: algorithm implementation discrepancy (11.9\%), language construct mismatch (66.4\%), and resource management inefficiency (21.7\%).
Third, inference-time prompt strategies offer limited relief. Few-shot prompting provides the most consistent yet modest efficiency gains, suggesting that improving translation efficiency likely requires more than prompt-level intervention. Taken together, our findings establish \textsc{Trace} as the first benchmark for jointly evaluating correctness and efficiency in LLM-based code translation, providing a principled foundation for future research.


This paper makes the following contributions: 1)\enspace\textbf{Dimension.} We are the first to identify execution efficiency as a critical dimension in LLM-based code translation. 
2)\enspace\textbf{Benchmark.} We present \textsc{Trace}, the first benchmark designed to expose efficiency issues in LLM-translated code. 
3)\enspace\textbf{Analysis.} We conduct a large-scale empirical and diagnostic study of 28 representative LLMs on \textsc{Trace}, revealing the misalignment between correctness and efficiency and deriving a code translation-specific taxonomy of inefficiencies.

\section{Related Work}

\begin{figure*}[t!]
\centering
\includegraphics[width=0.925\textwidth]{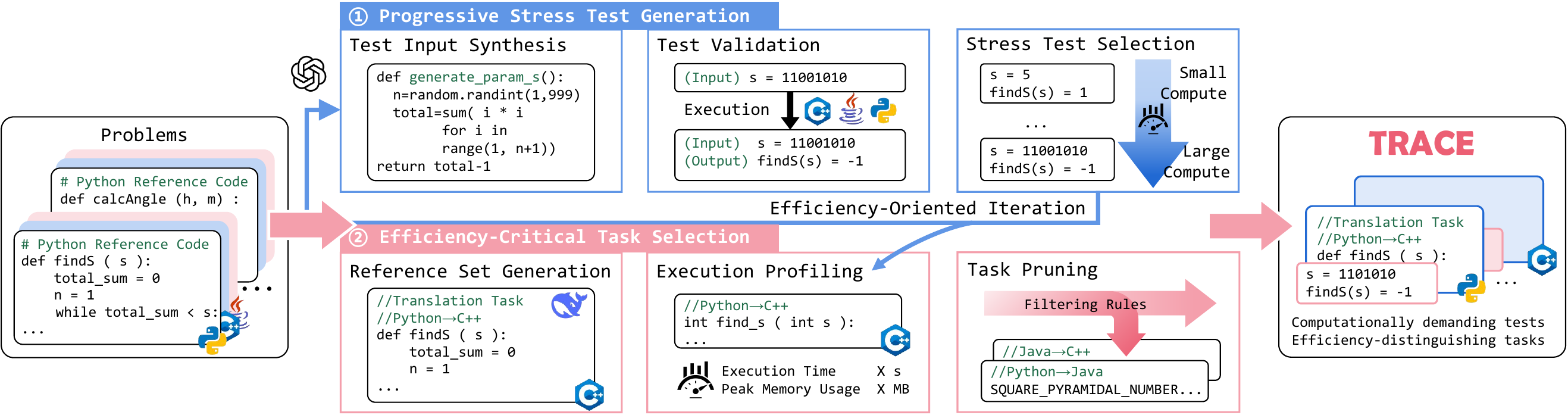} 
\caption{An overview of \textsc{Trace}'s LLM-driven two-stage construction pipeline.}
\label{fig:method_overview}
\end{figure*}

\textbf{Automatic Code Translation.}  
Automatic code translation has progressed from early rule-based~\cite{ct-c2rust, ct-cxgo} and statistical methods~\cite{ct-karaivanov2014phrase, ct-nguyen2016mapping} to neural approaches such as the Transcoder series~\cite{lachaux2020unsupervised, roziere2021leveraging, szafraniec2022code}. With the help of LLMs, research has mainly pursued correctness through fine-tuning~\cite{he2025execoder, liu2023syntax}, reasoning~\cite{macedo2024intertrans, nitin2024spectra}, and self-repair~\cite{yang2024exploring, pan2024lost}. Meanwhile, benchmarks such as Avatar~\cite{ahmad2021avatar}, G-TransEval~\cite{jiao2023evaluation}, PolyHumanEval~\cite{tao2024unraveling}, and ClassEval-T~\cite{xue2025classeval} have progressively expanded coverage across both programming languages and evaluation granularities. However, current work remains centered on functional correctness and leaves efficiency underexplored.

\textbf{Efficiency of LLM-generated Code.}  
Recent work has begun to study the execution efficiency of LLM-generated code. Benchmarks such as EffiBench~\cite{huang2024effibench} and EffiBench-X~\cite{qing2025effibench} construct efficiency-critical tasks, EvalPerf~\cite{liu2024evaluating} proposes stress-based differential performance evaluation to reveal efficiency differences under stress-test inputs, and Mercury~\cite{du2024mercury} formalizes the Beyond score, a runtime-percentile-weighted metric for finer-grained performance assessment. On the optimization side, methods such as PIE~\cite{shypula2023learning}, SwiftCoder~\cite{huang2024swiftcoder}, and Afterburner~\cite{du2025afterburner} adapt or fine-tune LLMs to generate more efficient code.

\textsc{Trace} connects and extends these two lines of research. First, it extends code translation evaluation beyond 
correctness by treating execution efficiency as an explicit quality dimension. Second, it studies efficiency in a \textit{cross-language} setting that differs from prior work on code generation~\cite{huang2024effibench, qing2025effibench, liu2024evaluating, du2024mercury}. Existing studies ask whether LLM-generated code is efficient for a given task within a single language. By contrast, \textsc{Trace} examines whether LLM-translated code derived from the same source program preserves efficiency-critical semantics in the target language. This setting exposes translation-specific efficiency failures that prior work does not capture, such as mapping an efficient data structure or idiom to a suboptimal target-language counterpart.

\section{Benchmark Construction}

\textbf{Terminology Formulation.} A problem ($P$) is a canonical programming challenge. For each problem, a reference code ($p$) is a functionally correct solution. A code translation task (task) is an instance of translating a specific reference code $p$ from a source language $l_s$ to a target language $l_t$, denoted as $p_{l_s} \rightarrow p_{l_t}$. This work focuses on two efficiency-critical indicators: execution time (ET) and peak memory usage (PM).

\textbf{Methodology Overview.} As presented in Figure~\ref{fig:method_overview}, \textsc{Trace} is constructed through an LLM-driven two-stage pipeline. The construction begins with established programming challenges. The first stage, \textit{Progressive Stress Test Generation}, iteratively synthesizes, validates, and selects stress tests through efficiency-oriented iteration, aiming to amplify latent efficiency disparities among translations. 
The second stage, \textit{Efficiency-Critical Task Selection}, applies rigorous filtering rules to isolate tasks that meaningfully differentiate efficiency.

\subsection{Problem Collection} 
\textsc{Trace} uses problems from the \textsc{Transcoder-Test}~\cite{lachaux2020unsupervised} benchmark, a widely used code translation benchmark containing diverse programming challenges. We select this foundation for two reasons: 1) It provides a high-quality set of recognized problems, ensuring relevance with prior work; 2) More importantly, it directly highlights the overlooked efficiency dimension within tasks that were previously evaluated only for correctness. We adopt the latest version of \textsc{Transcoder-Test}~\cite{yang2024exploring}, which contains 568 problems, each paired with one ground-truth implementation in C++, Java, and Python, along with 10 default correctness tests.

\subsection{Progressive Stress Test Generation}

We design an iterative stress test generation strategy, as detailed in Algorithm~\ref{alg:algorithm_stcg}. This strategy progressively generates \textit{stress tests} (tests demanding substantial computational resources during execution) to magnify hidden efficiency disparities.

\textbf{Test Input Synthesis.}
We first synthesize large-scale test inputs that sufficiently exercise the runtime behavior of the translated code. However, directly generating such inputs with an LLM is often impractical, because their explicit representation may exceed the LLM context or output length limits. We therefore utilize the \textit{synthesis-based} approach (Lines 5-7)~\cite{liu2024evaluating, shi2026codehacker, wang2025codecontestsplus}, where the LLM ($\mathcal{M}$) is prompted to generate multiple test input \textit{synthesizers} ($\mathcal{S}$) based on the ground-truth reference code. Each synthesizer is an independent Python program that programmatically outputs a test input when executed. To guide the LLM toward generating more resource-intensive inputs, we enrich the base prompt (Line 6) with synthesizer examples selected from the previous iteration. This enriched prompt ($\mathcal{E}$, Figure~\ref{figure:prompt_stress_test_generation}) incorporates: 1) two of the most effective synthesizers discovered so far; 2) the execution profiles of their outputs; and 3) chain-of-thought instructions. These modules jointly steer the LLM toward synthesizers that yield increasingly demanding test inputs.


\begin{algorithm}[t]
\footnotesize
\caption{Progressive Stress Test Generation}
\label{alg:algorithm_stcg}
\begin{algorithmic}[1]
    \STATE \textbf{FUNCTION} GenerateStressTests($P, \mathcal{M}, I_{max}, K$)
    \STATE \hspace{\algorithmicindent} $\mathcal{P} \leftarrow \text{GetBasePrompt}(P)$
    \STATE \hspace{\algorithmicindent} $\mathcal{T}_{s}, \mathcal{E} \leftarrow \emptyset, \emptyset$
    \STATE \hspace{\algorithmicindent} \textbf{for} $i \leftarrow 1$ \textbf{to} $I_{max}$ \textbf{do}
    \STATE \textcolor{DarkGreen}{\hspace{\algorithmicindent}\hspace{\algorithmicindent} // --- Stage 1.1: Test Input Synthesis ---}
    \STATE \hspace{\algorithmicindent}\hspace{\algorithmicindent} $\mathcal{P}_{iter} \leftarrow \text{EnrichPrompt}(\mathcal{P}, \mathcal{E})$
    \STATE \hspace{\algorithmicindent}\hspace{\algorithmicindent} $\mathcal{S} \leftarrow \mathcal{M}(\mathcal{P}_{iter})$

    \STATE \textcolor{DarkGreen}{\hspace{\algorithmicindent}\hspace{\algorithmicindent} // --- Stage 1.2: Test Validation ---}
    \STATE \hspace{\algorithmicindent}\hspace{\algorithmicindent} $\mathcal{V} \leftarrow \text{ExecuteSynthesizers}(\mathcal{S})$
    \STATE \hspace{\algorithmicindent}\hspace{\algorithmicindent} $\mathcal{T}_{cand} \leftarrow \text{Validate}(\mathcal{V})$
    
    \STATE \textcolor{DarkGreen}{\hspace{\algorithmicindent}\hspace{\algorithmicindent} // --- Stage 1.3: Stress Test Selection ---}
    \STATE \hspace{\algorithmicindent}\hspace{\algorithmicindent} $\mathcal{T}_{time}$ $\leftarrow$ SelectTopK($\mathcal{T}_{cand}$, $K$, \texttt{time})
    \STATE \hspace{\algorithmicindent}\hspace{\algorithmicindent} $\mathcal{T}_{mem}$ $\leftarrow$ SelectTopK($\mathcal{T}_{cand} \setminus \mathcal{T}_{time}$, $K$, \texttt{mem})
    \STATE \hspace{\algorithmicindent}\hspace{\algorithmicindent} $\mathcal{T}_{iter} \leftarrow \mathcal{T}_{time} \cup \mathcal{T}_{mem}$
    
    \STATE \textcolor{DarkGreen}{\hspace{\algorithmicindent}\hspace{\algorithmicindent} // --- Stage 1.4: Efficiency-Oriented Iteration ---}
    \STATE \hspace{\algorithmicindent}\hspace{\algorithmicindent} $\mathcal{T}_{s, \text{prev}} \leftarrow \mathcal{T}_{s}$
    \STATE \hspace{\algorithmicindent}\hspace{\algorithmicindent} $\mathcal{T}_{s} \leftarrow \text{UpdateGlobalPool}(\mathcal{T}_{s}, \mathcal{T}_{iter}, K)$
    \STATE \hspace{\algorithmicindent}\hspace{\algorithmicindent} \textbf{if} $\mathcal{T}_{s} = \mathcal{T}_{s, \text{prev}}$ \textbf{then} \textbf{break}
    \STATE \hspace{\algorithmicindent}\hspace{\algorithmicindent} \textbf{end if}
    \STATE \hspace{\algorithmicindent}\hspace{\algorithmicindent} $\mathcal{E} \leftarrow \text{ExtractExamples}(\mathcal{T}_s)$
    \STATE \hspace{\algorithmicindent} \textbf{end for}
    \STATE \hspace{\algorithmicindent} \textbf{return} $\mathcal{T}_s$
\end{algorithmic}
\end{algorithm}

\textbf{Test Validation.}
After executing the synthesizers, the generated test inputs are rigorously validated to ensure soundness and cross-language consistency (Lines 8-10). 
A test input is retained only if it satisfies two mandatory runtime checks: 1) Execution Integrity, where it must execute successfully across ground-truth reference code (C++, Java, and Python) without runtime exceptions; and 2) Output Consistency, where it must produce identical outputs across reference code. This step eliminates test inputs susceptible to language-specific behaviors (e.g., integer overflow in C++ versus Python's arbitrary-precision integers), establishing reliable ground-truths for validated candidate tests ($\mathcal{T}_{cand}$).

\textbf{Stress Test Selection.}
From the validated candidates $\mathcal{T}_{cand}$, we derive the stress test set $\mathcal{T}_{iter}$ for the current iteration (Lines 11-14). 
The targeted runtime indicators, ET and PM, are measured across reference code in C++, Java, and Python. 
For each indicator, we apply the \textit{Borda Count} method~\cite{emerson2013original} to rank the tests. A test is ranked within each language, and its overall score is the sum of its three language-specific ranks. A lower Borda score thus reflects tests that consistently consume more resources across languages. Finally, for both ET and PM, we select the top-$K$ tests with the lowest Borda scores to construct the stress test set $\mathcal{T}_{iter}$.

\textbf{Efficiency-Oriented Iteration.}
\label{subsubsection:efficiency_oriented_iteration}
The stress test generation proceeds as a feedback-driven loop to progressively discover computationally demanding tests (Lines 15-20). A global pool ($\mathcal{T}_s$) is maintained to store the most stressful tests identified so far. After each iteration, the newly selected tests ($\mathcal{T}_{iter}$) are merged into this pool. The pool is then re-ranked and pruned to retain only the overall top-$K$ tests for both ET and PM (Line 17). If the updated pool $\mathcal{T}_{s}$ remains unchanged, the loop terminates early, indicating convergence. Synthesizers producing these top-performing tests are collected as guiding examples ($\mathcal{E}$) to enrich the synthesis prompt (Line 20) for the next iteration.

\subsection{Efficiency-Critical Task Selection}
Not all tasks are suitable for evaluating efficiency. For instance, a trivial translation of \textit{def add(a, b): return a+b} is unlikely to produce variants with notable performance differences. To address this, we apply rigorous filtering to retain only tasks that meaningfully expose efficiency disparities.

For each task, we generate a diverse pool of reference translations by sampling from a representative set of LLMs. 
Each candidate translation is validated against the default tests from \textsc{Transcoder-Test} to ensure functional correctness. 
The correct translations are retained and executed on the newly generated stress tests to measure ET and PM. 
The resulting profiled translations form the task-level \textit{efficiency reference set}.

Based on the efficiency reference set, we apply three filtering criteria:  
1) Feasibility. A task is discarded if no evaluated LLM produces a functionally correct translation, as no valid samples remain for efficiency profiling.  
2) Impactfulness. A task with trivial runtime cost is removed, since efficiency is not a practical concern. Concretely, at least one reference translation must exceed a predefined execution time threshold ($\epsilon_{T}$) or peak memory threshold ($\epsilon_{M}$) under stress tests.  
3) Diversity. A task is excluded if correct translations exhibit negligible performance variance. This is measured by requiring the coefficient of variation ($\text{CV} = \text{std} / \text{mean}$) for either ET or PM to exceed a threshold ($\epsilon_{D}$), ensuring that the task can meaningfully expose efficiency disparities.

\subsection{Benchmark Implementation}
\label{section:implementation}


For the \textit{Progressive Stress Test Generation} stage, we employ GPT-4o (temperature $=0.8$) to generate stress test synthesizers. 
We prompt GPT-4o to produce 3 distinct synthesizers for each ground-truth reference code across C++, Java, and Python, resulting in 9 synthesizers per problem. 
Each synthesizer is executed 3 times, yielding up to 27 candidate test inputs per iteration. 
The iterative loop (Algorithm~\ref{alg:algorithm_stcg}) is capped at a maximum of $I_{\max}=5$ iterations. 
In each iteration, we select the top $K=5$ tests by ET and another $K=5$ by PM, giving 10 stress tests per iteration. 
Aggregating across iterations, the final stress test suite for each problem contains the 10 most computationally demanding tests.

For the \textit{Efficiency-Critical Task Selection} stage, we first construct a diverse pool of reference translations using 28 representative LLMs (see Section~\ref{section:experiment_setup}), decoded with sampling (temperature $=0.8$). To select efficiency-critical tasks, we apply empirically derived thresholds: impactfulness ($\epsilon_{T}=0.01$ seconds, $\epsilon_{M}=1.5$ MB) and diversity ($\epsilon_{D}=0.05$). 
Following prior work~\cite{liu2024evaluating}, these thresholds are estimated from the average resource consumption of a simple \textit{Hello World} program across C++, Java, and Python, thereby excluding tasks with negligible runtime fluctuations.


Further benchmark construction details on the model choice and threshold selection are provided in Appendix~\ref{subsection:pipeline_design_ablation}.

\subsection{Benchmark Statistics}


\begin{table}[t!]
\centering
\footnotesize
\caption{Key statistics of the \textsc{Trace} benchmark.}
\label{table:stats_of_tracy}
\begin{tabularx}{.95\linewidth}{@{}X r@{}}
\toprule
\textbf{Benchmark Profile} & \textbf{Detail} \\
\midrule
\textbf{Total Problems} & 357 \\
\textbf{Total Tasks} & 1,000 \\
\midrule
\textbf{Default Tests (Per Task)} & 10 \\
\quad Execution Time (s) & 0.09 \\
\quad Peak Memory (MB) & 24.35 \\
\midrule
\textbf{Stress Tests (Per Task)} & 10 \\
\quad Execution Time (s) & 0.80 (8.9$\times$) \\
\quad Peak Memory (MB) & 83.09 (3.4$\times$) \\
\midrule
\textbf{Efficiency Reference Set (Per Task)} & 23.03 \\
\quad\quad Performance Clusters & 3.94 \\
\bottomrule
\end{tabularx}
\end{table}

Table~\ref{table:stats_of_tracy} summarizes the key characteristics of \textsc{Trace}, and Appendix Table~\ref{table:benchmark_comparison} further compares \textsc{Trace} with existing code translation benchmarks.

Overall, \textsc{Trace} contains 357 problems, yielding a total of 1,000 efficiency-critical tasks across six translation directions. Specifically, the distribution covers {C++$\to$Java (250)}, {Java$\to$C++ (127)}, {Java$\to$Python (148)}, {Python$\to$Java (236)}, {C++$\to$Python (141)}, and {Python$\to$C++ (98)} tasks.
Each task retains 10 default tests from \textsc{Transcoder-Test} and is further equipped with 10 newly generated stress tests. 
On average, \textsc{Trace}'s stress tests increase ET and PM by $8.9\times$ and $3.4\times$, respectively, effectively exposing inefficiencies that otherwise remain invisible under small-scale tests. 
Each task is associated with an efficiency reference set averaging 23 correct translations. Clustering their ET/PM profiles with HDBSCAN~\cite{mcinnes2017hdbscan} yields 3.94 distinct performance clusters per task, revealing substantial heterogeneity in the efficiency behaviors of translations and enabling fine-grained evaluation.

Compared with existing benchmarks, \textsc{Trace} introduces two key innovations. 
First, it explicitly integrates stress tests to exercise computational complexity, thereby revealing inefficiency overlooked by small-scale tests. 
Second, it selectively retains efficiency-critical tasks via rigorous filtering, ensuring that every included task provides meaningful variation in execution efficiency.

\begin{table*}[ht]
\centering
\footnotesize
\caption{Overall performance of representative proprietary and open-source LLMs on \textsc{Trace}.}

\setlength{\tabcolsep}{3.20pt}
\renewcommand{\arraystretch}{0.95}
\label{table:overall_performance}
\begin{tabular}{
l
>{\columncolor{gray!10}}c 
>{\columncolor{gray!10}}c 
>{\columncolor{gray!10}}c 
>{\columncolor{gray!10}}c 
>{\columncolor{gray!10}}c 
*{6}{cc}
}
\toprule
\multirow{2}{*}{\textbf{Model}}
& \multicolumn{5}{>{\columncolor{gray!10}}c}{\textbf{Summary}}
& \multicolumn{2}{c}{\textbf{C++$\rightarrow$Ja}}
& \multicolumn{2}{c}{\textbf{C++$\rightarrow$Py}}
& \multicolumn{2}{c}{\textbf{Ja$\rightarrow$C++}}
& \multicolumn{2}{c}{\textbf{Ja$\rightarrow$Py}}
& \multicolumn{2}{c}{\textbf{Py$\rightarrow$C++}}
& \multicolumn{2}{c}{\textbf{Py$\rightarrow$Ja}} \\
\cmidrule(lr){2-6}\cmidrule(lr){7-8}\cmidrule(lr){9-10}\cmidrule(lr){11-12}\cmidrule(lr){13-14}\cmidrule(lr){15-16}\cmidrule(lr){17-18}
& \textbf{Pass} & $\boldsymbol{B_T}$ & $\boldsymbol{B_M}$ & $\boldsymbol{B_{T}^P}$ & $\boldsymbol{B_{M}^P}$
& $\boldsymbol{B_T}$ & $\boldsymbol{B_M}$
& $\boldsymbol{B_T}$ & $\boldsymbol{B_M}$
& $\boldsymbol{B_T}$ & $\boldsymbol{B_M}$
& $\boldsymbol{B_T}$ & $\boldsymbol{B_M}$
& $\boldsymbol{B_T}$ & $\boldsymbol{B_M}$
& $\boldsymbol{B_T}$ & $\boldsymbol{B_M}$ \\
\midrule
\multicolumn{18}{c}{\textit{Proprietary LLMs}} \\
\midrule
Claude-4-Think & \textbf{95.5} & 49.6 & 50.5 & 51.9 & 52.9 & 37.1 & 54.0 & 61.3 & 48.4 & 58.8 & 50.7 & 65.4 & 46.9 & 52.0 & 44.4 & 39.9 & 52.6 \\
Claude-4       & 93.4 & 48.8 & 45.5 & 52.3 & 48.7 & 66.8 & 62.9 & 25.7 & 13.2 & 45.5 & 56.6 & 27.0 & 16.1 & 43.4 & 47.5 & 61.1 & 58.1 \\
DS-Reasoner    & 72.7 & 39.3 & 28.1 & 54.1 & 38.7 & 34.6 & 28.4 & 45.3 & 36.0 & 31.4 & 23.9 & 53.8 & 36.9 & 32.8 & 19.3 & 38.5 & 23.5 \\
DS-V3.2        & 91.9 & \textbf{50.7} & 44.4 & 55.1 & 48.3 & 67.2 & 63.3 & 30.0 & 16.3 & 49.4 & 46.5 & 33.0 & 14.9 & 30.4 & 31.7 & 65.6 & 63.7 \\
Gemini-Pro     & 62.5 & 37.9 & 24.5 & \textbf{60.7} & 39.3 & 35.9 & 20.7 & 44.8 & 34.7 & 29.3 & 22.2 & 47.6 & 35.8 & 27.3 & 10.0 & 38.9 & 22.8 \\
Gemini-Flash   & 62.5 & 37.6 & 25.5 & 60.2 & 40.8 & 37.2 & 26.2 & 49.8 & 37.0 & 33.2 & 22.6 & 51.3 & 35.0 & 18.1 & 7.9  & 32.8 & 20.9 \\
O3             & 85.1 & 50.2 & 34.8 & 58.9 & 40.8 & 53.5 & 34.3 & 63.6 & 49.0 & 39.9 & 25.1 & 61.0 & 46.5 & 26.9 & 17.2 & 46.9 & 31.8 \\
O3-Mini        & 91.0 & 42.9 & \textbf{50.7} & 47.2 & \textbf{55.8} & 40.7 & 52.3 & 37.4 & 63.5 & 48.9 & 47.3 & 58.4 & 46.1 & 33.0 & 32.2 & 39.7 & 54.0 \\
\midrule
\multicolumn{18}{c}{\textit{Open-Source LLMs}} \\
\midrule
CL-7B-Inst     & 75.5 & 44.2 & 30.6 & 58.5 & 40.6 & 39.1 & 28.6 & 53.5 & 39.2 & 43.1 & 33.2 & 53.7 & 38.0 & 34.2 & 15.8 & 42.7 & 27.7 \\
CL-13B-Inst    & 76.7 & 45.3 & 31.3 & 59.1 & 40.8 & 40.7 & 26.0 & 53.0 & 39.8 & 43.7 & 34.9 & 56.5 & 39.2 & 32.8 & 26.5 & 44.7 & 27.0 \\
CL-34B-Inst    & 69.3 & 42.2 & 30.3 & \textbf{60.9} & 43.7 & 36.6 & 24.1 & 54.8 & 43.4 & 30.4 & 29.2 & 56.3 & 40.1 & 30.0 & 20.6 & 43.3 & 27.3 \\
DSC-6.7B-Inst  & 86.2 & 51.0 & 35.3 & 59.1 & 40.9 & 50.9 & 33.7 & 58.4 & 44.9 & 42.4 & 34.2 & 62.1 & 43.6 & 36.4 & 22.5 & 50.3 & 31.9 \\
DSC-33B-Inst   & 89.8 & 52.2 & 36.9 & 58.2 & 41.1 & 54.8 & 36.5 & 60.6 & 46.9 & 41.3 & 27.2 & 64.3 & 46.1 & 40.1 & 30.5 & 47.9 & 33.6 \\
QC-7B-Inst     & 90.1 & 51.4 & 36.9 & 57.1 & 41.0 & 42.5 & 33.0 & 64.3 & 44.8 & 48.5 & 39.2 & 64.9 & 47.4 & 43.2 & 28.3 & 49.6 & 32.1 \\
QC-14B-Inst    & \textbf{91.6} & \textbf{54.8} & 39.0 & 59.8 & 42.6 & 54.3 & 36.5 & 63.3 & 46.5 & 46.2 & 37.4 & 65.8 & 46.0 & 42.4 & 29.4 & 53.0 & 37.7 \\
QC-32B-Inst    & 90.7 & 48.7 & \textbf{42.9} & 53.7 & \textbf{47.3} & 56.2 & 37.7 & 63.8 & 44.7 & 50.3 & 40.6 & 33.8 & 68.2 & 26.7 & 31.7 & 49.3 & 37.3 \\
\midrule
\textbf{Average} & 84.9 & 45.8 & 37.7 & 54.0 & 43.6 & 45.8 & 37.7 & 52.6 & 39.7 & 41.4 & 34.6 & 54.2 & 39.2 & 35.7 & 27.5 & 45.5 & 37.3 \\

\bottomrule
\end{tabular}
\end{table*}

\section{Experiment Setup}
\label{section:experiment_setup}

We first outline models, metrics, and the evaluation protocol used in our study.

\textbf{Models.}
We evaluate 28 representative LLMs, covering both leading proprietary and prominent open-source code LLMs. For proprietary models, we include OpenAI (GPT-4o and O3 series), Anthropic (Claude-4 and 3.5 series), Google (Gemini-2.5 series), and DeepSeek. For open-source models, we assess CodeLlama-hf (CL)~\cite{roziere2023code}, DeepSeek-Coder (DSC)~\cite{guo2024deepseek}, and Qwen2.5-Coder (QC) series~\cite{hui2024qwen2}, including both base and instruction-tuned (Inst) models. Table~\ref{table:model_list} summarizes the model aliases and details.

\textbf{Metrics.}
We evaluate model performance along two complementary dimensions: functional correctness and execution efficiency. 

For correctness, we use \textit{Pass Rate} (Pass), defined as the percentage of tasks whose translations pass all tests.

For efficiency, we adopt the \textit{Beyond} score~\cite{du2024mercury}. For a given efficiency indicator $X$ (e.g., execution time), the score evaluates each candidate translation $c$ against the task-level efficiency reference set $\mathcal{R}$. Specifically, let $\mathcal{R}_X = \{X(r) : r \in \mathcal{R}\}$ denote the task-specific reference values for indicator $X$, and let $x_c = X(c)$ denote the candidate’s measured value. For brevity, let $r_X^{\min}=\min \mathcal{R}_X$ and $r_X^{\max}=\max \mathcal{R}_X$. Beyond is defined as:

{\small
\begin{equation*}
B_X(c)=
\frac{
r_X^{\max}-\operatorname{clip}\!\left(x_c, r_X^{\min}, r_X^{\max}\right)
}{
r_X^{\max}-r_X^{\min}
}
\times 100\%.
\end{equation*}
}

where $\operatorname{clip}(a,\ell,u)$ truncates $a$ to $[\ell,u]$. Intuitively, $B_X(c)$ reflects the normalized efficiency of candidate $c$ relative to the task-specific reference range, where a higher score indicates better efficiency. In this work, we consider two indicators, ET and PM, and compute Beyond separately for each. We further report two variants: $B_X$, averaged over all translations with incorrect outputs assigned a score of 0, and $B_X^P$, computed only over correct translations to isolate efficiency under correctness.

\textbf{Model Evaluation Protocol.}  
For each task, the model generates a single translation using greedy decoding (temperature $=0.0$) under a zero-shot prompt (Appendix~\ref{section:translation_prompt}). 
Functional correctness is verified with the default tests from \textsc{Transcoder-Test}. Translations that fail to compile, trigger runtime errors, or exceed resource limits (180\,s or 4096\,MB) are marked as incorrect and excluded from efficiency evaluation.  
Correct translations are then executed on stress tests to measure ET and PM, with each metric reported as the arithmetic mean of five runs. Further experimental details are provided in Appendix~\ref{subsection:experiment_environment}.


\section{Evaluation}

We conduct extensive evaluation through three lenses: 1) an end-to-end model performance comparison, 2) a taxonomy-driven characterization of inefficient translations, and 3) an exploration of inference-time efficiency improvement strategies.

\subsection{Overall Performance Landscape}
Table~\ref{table:overall_performance} reports the performance of a representative set of LLMs to illustrate our key findings. The complete evaluation is reported in Table~\ref{table:granular_performance_comparison}.


\begin{table*}[t]
\centering
\footnotesize
\setlength{\tabcolsep}{5pt}
\caption{Taxonomy of inefficiency patterns observed in LLM-based code translation.}
\label{table:inefficiency_taxonomy_compact}
\renewcommand{\arraystretch}{0.925}
\begin{tabularx}{.95\linewidth}{@{}X r@{}}
\toprule
\textbf{Category} & \textbf{Pct.(\%)} \\
\midrule
\cellcolor{gray!10} \textbf{1. Algorithm Implementation Discrepancy} & \textbf{11.93} \\
\quad \textbf{1.1 Asymptotic Complexity Degradation} & \textbf{3.36} \\
\quad The translation adopts an alternative algorithm with worse asymptotic complexity than the source code. & \\
\quad \textbf{1.2 Omission of Target-Language Idioms} & \textbf{8.56} \\
\quad The translation ignores efficient idioms or optimized library routines available in the target language. & \\
\cellcolor{gray!10} Example: C++ in–place reversal \textbf{std::reverse(s.begin(), s.end())} ($O(N)$) is translated into Python repeated concatenation \textbf{res=\texttt{""}; for c in s: res=c+res} ($O(N^2)$), instead of the idiomatic target \textbf{s[::-1]}. & \\
\cmidrule(lr){1-2}

\cellcolor{gray!10} \textbf{2. Language Construct Mismatch} & \textbf{66.36} \\
\quad \textbf{2.1 Suboptimal Data Structure Selection} & \textbf{20.49} \\
\quad The adopted data structure is asymptotically or practically inferior for the required operations. & \\
\quad \textbf{2.2 Inefficient API/Pattern Usage} & \textbf{45.87} \\
\quad The translation uses non-optimal APIs or coding patterns and introduces unnecessary runtime costs. & \\
\cellcolor{gray!10} Example: Java \textbf{HashMap} (amortized $O(1)$) is mapped to C++ \textbf{std::map} ($O(\log N)$) instead of \textbf{std::unordered\_map}, which becomes inefficient when handling frequent lookups or insertions.  & \\
\cmidrule(lr){1-2}

\cellcolor{gray!10} \textbf{3. Resource Management Inefficiency} & \textbf{21.71} \\
\quad \textbf{3.1 Inefficient Data Representation} & \textbf{17.13} \\
\quad Primitive or compact source representations are replaced with memory-intensive object-based equivalents. & \\
\quad \textbf{3.2 Non-Idiomatic Memory Management} & \textbf{4.59} \\
\quad The translation adopts manual or unsafe memory practices instead of modern, idiomatic abstractions. & \\
\cellcolor{gray!10} Example: C++ \textbf{long} is translated into Java \textbf{BigInteger}, causing excessive allocation and memory footprint; when the value range permits, the optimal target is \textbf{long}. & \\
\bottomrule
\end{tabularx}
\end{table*}

\begin{figure}[t!]
\centering
\includegraphics[width=.925\linewidth]{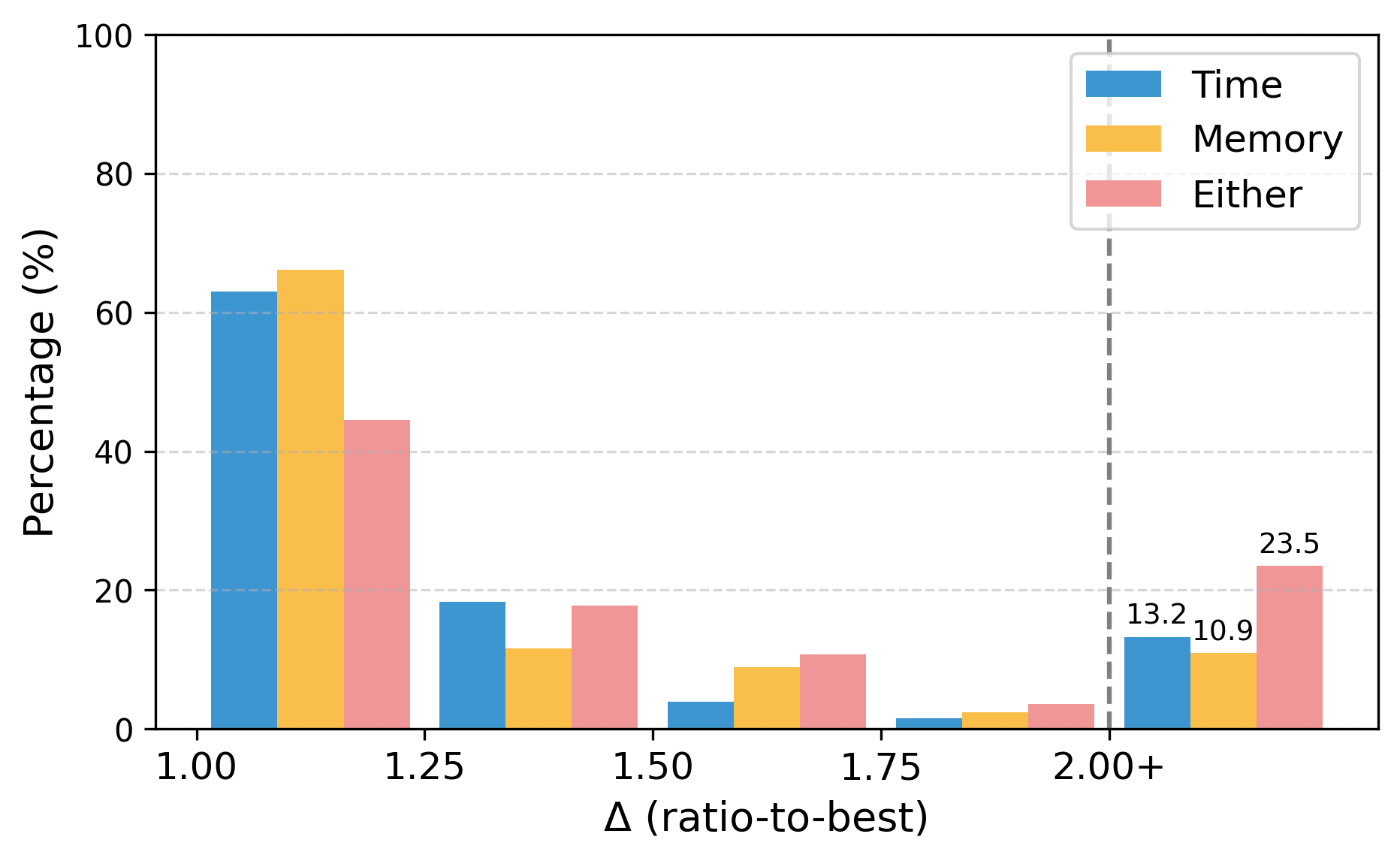} 
\caption{Execution profile distribution of translations.}
\label{figure:execution_distribution}
\end{figure}

\textbf{\textit{Correctness is not a proxy for efficiency.}}  
From Table~\ref{table:overall_performance}, we find that higher correctness does not necessarily lead to higher efficiency. For instance, Claude-4-Think attains the highest Pass rate (95.5) but only moderate time efficiency ($B_T$: 49.6), outperformed by smaller open-source LLMs such as QC-14B-Inst ($B_T$: 54.8). Conversely, CL-34B-Inst achieves the lowest correctness among open-source models (69.3), yet it delivers the most time-efficient solutions for its correct translations ($B_T^P$: 60.9).

To examine the relationship between correctness and efficiency, we conduct correlation analysis (Appendix~\ref{section:relationship_correctness_and_efficiency}). Across LLMs, correctness shows a moderate negative association with time efficiency (Pearson $r=-0.54$, Spearman $\rho=-0.61$) and a moderate positive association with memory efficiency ($r=0.57$, $\rho=0.70$). Restricting the analysis to high-performing models (Pass rate $\geq 85\%$), evaluated on their 613 jointly solved tasks, yields similar trends ($r=-0.51$ for ET, $r=0.64$ for PM). OLS regression further confirms that correctness explains only a limited portion of efficiency variance ($R^2=0.29$ for ET, $R^2=0.33$ for PM).

Our findings indicate that correctness and efficiency are distinct dimensions of performance, with correctness alone insufficient to explain the observed efficiency variation.

\textbf{\textit{Reasoning and scaling do not reliably improve efficiency.}} 
Across proprietary and open-source models, neither reasoning-oriented variants nor larger models consistently improve efficiency. Reasoning models, trained with intermediate reasoning supervision, do not always outperform their standard counterparts: for instance, while Claude-4-Think achieves slightly higher time efficiency than Claude-4 ($B_T$: 49.6 vs.\ 48.8), the reverse is observed in the DeepSeek family, where the general-purpose DS-V3.2 surpasses DS-Reasoner in terms of both time ($B_T$: 50.7 vs.\ 39.3) and memory efficiency ($B_M$: 44.4 vs.\ 28.1). Second, scaling in model size likewise provides no monotonic improvements. Larger models generally exhibit stronger capabilities in code understanding and generation, yet these advantages often fail to yield corresponding efficiency gains. For instance, QC-14B-Inst is more time-efficient than its 32B counterpart ($B_T$: 54.8 vs. 48.7), and within the CodeLlama series, the 34B variant ($B_T$: 42.2) is surpassed by the 7B (44.2) and 13B (45.3) models.

These results suggest that existing reasoning and model scaling practices remain primarily correctness-driven, leaving efficiency as a secondary and unresolved challenge.

\textbf{\textit{Efficiency exhibits strong directional asymmetry across languages.}}  
We observe a clear asymmetry across translation directions. For Java$\leftrightarrow$Python, Pass rates are relatively close (89.3 vs.\ 85.2, Table~\ref{table:granular_performance_comparison}). However, their time efficiency differs sharply ($B_T$: 54.2 vs.\ 45.5), and this gap remains even when considering only correct translations ($B_T^P$: 61.2 vs.\ 53.5). These results suggest that LLMs internalize language-specific knowledge unevenly, producing more efficient translations into Python than into Java. This asymmetry shows that efficiency is shaped not only by model capacity but also by structural properties of programming languages, highlighting the challenge of building efficiency-aware multilingual models.

\subsection{Inefficiency Profile and Taxonomy}
We analyze the distribution and root causes of translation inefficiencies. Figure~\ref{figure:execution_distribution} shows the execution profile distribution, where each translation’s ET/PM is measured relative to the most efficient within the task. Overall, 23.5\% of translations exhibit substantial inefficiency, with ET/PM exceeding twice that of the task-specific best. This shows that inefficiency is not an isolated anomaly but a recurrent phenomenon even in correct translations.

To capture the root causes of inefficiency, we examined 327 representative cases (one case per problem) with an inefficiency ratio of over 2.0. Two authors independently classified the cases and resolved disagreements through discussion. The resulting taxonomy is summarized in Table~\ref{table:inefficiency_taxonomy_compact}.

\begin{table}[t!]
\centering
\small
\renewcommand{\arraystretch}{0.9}
\caption{Statistics of manually classified inefficiency cases. 
$\Delta T$/$\Delta M$ denotes the ratio of ET/PM of each translation relative to the task-specific best.}
\label{table:root_case_category}
\setlength{\tabcolsep}{4pt}
\begin{tabular}{@{}lrrrr@{}}
\toprule
\multicolumn{1}{@{}l}{\textbf{Inefficiency Category}} 
& \multicolumn{2}{c}{$\boldsymbol{\Delta T}$}
& \multicolumn{2}{c}{$\boldsymbol{\Delta M}$} \\
\cmidrule(lr){2-3}\cmidrule(lr){4-5}
& \textbf{Avg.} &\textbf{Med.}& \textbf{Avg.} & \textbf{Med.} \\
\midrule
Algorithm Implementation   & 371.6 & 3.2 & 2.5 & 1.2 \\
Language Construct &   6.6 & 2.1 & 3.4 & 1.9 \\
Resource Management  &   2.6 & 1.1 & 5.9 & 5.0 \\
\bottomrule
\end{tabular}
\end{table}

The examination reveals three major inefficiency patterns in code translation.
1) Algorithm Implementation Discrepancy (11.93\%). We classify a case as this type when the target algorithm diverges in computational complexity from the source or fails to employ efficient target-language idioms. It comprises two subtypes: asymptotic complexity degradation and omission of target-language idioms.
2) Language Construct Mismatch (66.36\%). We classify a case as this type when the source specifies an explicit construct, yet the translation maps it to a correct but suboptimal alternative. It comprises two subtypes: suboptimal data structure selection and inefficient API/pattern usage.
3) Resource Management Inefficiency (21.71\%). We classify a case as this type when the source specifies no representation, yet the translation introduces heavyweight abstractions or overhead. It comprises two subtypes: inefficient data representation and non-idiomatic memory management.

Table~\ref{table:root_case_category} quantifies the severity of inefficiencies across categories. Algorithm implementation discrepancies are relatively rare but induce catastrophic slowdowns in ET (avg.~371.6$\times$, median~3.2$\times$). Language construct mismatches dominate in frequency and introduce substantial overheads in both ET (avg.~6.6$\times$, median~2.1$\times$) and PM (avg.~3.4$\times$, median~1.9$\times$). Resource management inefficiencies impose the largest PM penalties, with an average increase of 5.9$\times$ and a median of 5.0$\times$. 

Our analysis shows that inefficiencies are both widespread and patterned, spanning algorithmic, construct-level, and resource-level dimensions.

\begin{table}[t]
\centering
\footnotesize
\setlength{\tabcolsep}{5pt}
\renewcommand{\arraystretch}{0.925} 
\caption{Impact of prompt strategies on efficiency.}
\label{table:prompt_effect}
\begin{tabular}{lrrrrr}
\toprule
\textbf{Prompt Strategy} & \textbf{Pass} & $\boldsymbol{B_T}$ & $\boldsymbol{B_M}$ & $\boldsymbol{B_T^P}$ & $\boldsymbol{B_M^P}$ \\
\midrule
\multicolumn{6}{l}{\textbf{GPT-4o}} \\ \cmidrule(lr){1-6}
\quad zero-shot        & 89.5 & 48.8 & 42.2 & 54.5 & 47.2 \\
\quad perf-zero-shot   & 89.4 & 50.5 & 44.9 & \textbf{56.5} & \textbf{50.2} \\
\quad perf-few-shot    & \textbf{93.9} & \textbf{51.6} & \textbf{46.3} & 54.9 & {49.3} \\
\quad perf-self-refine & 89.2 & 49.2 & 44.0 & {55.2} & 49.4 \\
\midrule
\multicolumn{6}{l}{\textbf{DS-V3.2}} \\ \cmidrule(lr){1-6}
\quad zero-shot        & 91.9 & 50.7 & 44.4 & 55.1 & 48.3 \\
\quad perf-zero-shot   & 90.2 & 49.9 & 42.7 & 55.3 & 47.4 \\
\quad perf-few-shot    & \textbf{94.4} & \textbf{52.9} & \textbf{47.5} & \textbf{56.1} & \textbf{50.3} \\
\quad perf-self-refine & 91.7 & 51.4 & 43.9 & \textbf{56.1} & 47.8 \\
\midrule
\multicolumn{6}{l}{\textbf{Claude-4}} \\ \cmidrule(lr){1-6}
\quad zero-shot        & 93.4 & 48.8 & 45.5 & 52.3 & 48.7 \\
\quad perf-zero-shot   & 91.6 & 47.9 & 42.8 & 52.3 & 46.7 \\
\quad perf-few-shot    & \textbf{94.3} & 50.2 & \textbf{48.7} & 53.3 & \textbf{51.7} \\
\quad perf-self-refine & 91.9 & \textbf{52.3} & 45.9 & \textbf{56.9} & 49.9 \\
\bottomrule
\end{tabular}
\end{table}

\subsection{Impact of Prompt Strategies on Efficiency}

We further explore whether inference-time prompt strategies can improve code translation efficiency. Beyond the default \textit{zero-shot} baseline, we experiment with three performance-aware prompts (Appendix~\ref{section:translation_prompt}), inspired by recent code efficiency work~\cite{shypula2023learning, liu2024evaluating, huang2024effilearner}:  
1) \textit{perf-zero-shot}, which adds an explicit system instruction on efficiency;  
2) \textit{perf-few-shot}, which provides two efficient translation examples; and  
3) \textit{perf-self-refine}, which refines the initial translation based on execution feedback.

Table~\ref{table:prompt_effect} summarizes the results. Overall, prompting yields modest and model-dependent gains. The most consistent improvements arise with perf-few-shot, which increases both correctness and efficiency (e.g., DS-V3.2, Pass: 94.4; $B_T$: 52.9, $B_M$: 47.5). Other strategies are mixed: perf-self-refine, for instance, improves Claude-4’s time efficiency ($B_T$: 48.8$\rightarrow$52.3) while slightly reducing correctness (93.4$\rightarrow$91.9). Taken together, the results suggest that prompt-based strategies can alleviate inefficiencies to a limited extent but fail to eliminate them, highlighting the need for more advanced methods to encode efficiency principles.


\section{Discussion}




\subsection{Validity of the Efficiency Reference Set}
\label{section:construct_validity}

A potential threat to construct validity lies in using an LLM-derived translation set (efficiency reference set) as the evaluation anchor. Since the Beyond score is defined relative to the observed reference spectrum, its validity depends on whether this set provides a sufficiently strong upper bound. If the reference set were weaker than human-optimized implementations, the resulting scores could overestimate the model performance.

To examine this risk, we conducted a human-baseline sanity check on 50 randomly sampled tasks (18 C++, 21 Java, and 11 Python). Two authors with over seven years of programming experience collaboratively implemented optimized solutions (\textit{Human}), which we compared against the fastest LLM-produced reference translation for each task (\textit{FastLLM}). Overall, \textit{Human} achieves an average execution time of 2.61s, compared with 2.63s for \textit{FastLLM} (a relative gap of about 0.8\%). The same pattern holds across languages: 0.50s vs. 0.53s in C++, 0.56s vs. 0.55s in Java, and 10.00s vs. 10.03s in Python. These results suggest that the reference set reaches a comparable efficiency level to human-optimized solutions, supporting its use as a reasonable evaluation anchor. Future work should extend this validation with larger samples and broader expert-written baselines.

\subsection{Towards Intrinsic Efficiency Awareness}

Our findings suggest that improving LLM-based code translation efficiency requires more than simple inference-time prompt engineering. This work provides two foundations for future research on efficiency-aware code translation. First, we offer a controlled benchmark, \textsc{Trace}, for evaluating whether new methods improve translation efficiency. Second, our diagnostic analysis identifies recurring inefficiency patterns. In particular, our taxonomy can guide the construction of targeted training data or optimization objectives. Future work can build on these signals through supervised fine-tuning or reinforcement learning to improve models' efficiency awareness in code translation.
\section{Conclusion}
This work foregrounds execution efficiency as a critical yet long-overlooked dimension in LLM-based code translation. 
We introduce \textsc{Trace}, a benchmark that explicitly exposes efficiency gaps beyond correctness through progressive stress test generation and efficiency-critical task selection. 
From an evaluation of 28 models, we find that correctness does not necessarily imply efficiency; inefficiencies are both prevalent and patterned; and inference-time prompt strategies deliver limited and model-dependent gains. Ultimately, our work highlights the open challenge of developing advanced methods that equip LLMs with stronger efficiency awareness for code translation.

\section*{Acknowledgments}
This work is supported by National Key Research and Development Program of China 2024YFE0204200, the National Natural Science Foundation of China under Grant No.~62372005 and 62402482. Jie M. Zhang is supported by the ITEA GreenCode Project under Project No.~23016 and the ITEA GENIUS Project under Project No.~23026.

\clearpage
\section*{Limitations}

The following limitations define the scope of our work and highlight avenues for future research.

\textbf{Scope of Efficiency.} 
Our study focuses on two core dimensions of efficiency: execution time and peak memory usage, following prior work on LLM-generated code efficiency~\cite{du2024mercury, huang2024effibench, qing2025effibench, liu2024evaluating, peng2025coffe}. 
While these metrics serve as fundamental indicators of execution efficiency, we acknowledge that they only partially capture real-world performance.  
Factors such as I/O latency, compilation overhead, and energy consumption may also substantially influence efficiency in practice.  
Future work should consider broader and domain-specific aspects to provide a more comprehensive view.

\textbf{Scope of Programming Languages.} 
Our study focuses on C++, Java, and Python, which consistently rank among the top three most popular languages in established popularity indices~\cite{pypl,tiobe}.
Concentrating on these languages provides strong representativeness and supports the significance of our findings. 
Meanwhile, we acknowledge that excluding other languages may pose risks to broader generalization. Extending \textsc{Trace} to additional programming languages would help evaluate the generalizability of our findings.

\textbf{Scope of Translation Scenario.}
The problems in \textsc{Trace} are drawn from \textsc{Transcoder-Test}, whose tasks were originally collected from GeeksForGeeks~\cite{geeksforgeeks}. 
Accordingly, \textsc{Trace} mainly studies method-level algorithmic and data-structure-oriented translation tasks. 
We acknowledge that this setting is cleaner than real-world translation scenarios, which often involve class-level or repository-level contexts~\cite{jimenez2023swe, ibrahimzada2025alphatrans, xue2025classeval}. Currently,
we focus on method-level translation because it provides a controlled setting for efficiency evaluation, and it serves as a core operational unit in higher-level translation workflows. For example, prior class- and repository-level approaches often decompose larger translation units into methods~\cite{xue2025classeval,ibrahimzada2025alphatrans}. As a result, inefficiencies introduced at the method level can carry over to higher-level settings. 
Future work should extend \textsc{Trace} to class-level and repository-level settings to cover more software translation scenarios.


\clearpage
\bibliography{reference}

\clearpage

\appendix

\section{Implementation Details}

\subsection{Pipeline Design Ablations}
\label{subsection:pipeline_design_ablation}

We provide details to justify key design choices in our benchmark construction pipeline.

\textbf{Model and Iteration Choice.} We performed a pilot study on a 10\% subset of \textsc{Transcoder-Test} (56 problems), comparing three widely used LLMs, \text{GPT-4o}, \text{Claude-4}, and \text{DS-V3.2}, for stress test generation. Table~\ref{table:pipeline_ablation_model_iteration} reports the average execution time (ET) and peak memory usage (PM) of the generated tests across iterations. For all three models, ET and PM increase substantially in the first five iterations, while the gains diminish beyond iteration 5. We therefore set the maximum number of iterations to $I_{\max}=5$. Among the compared models, \text{GPT-4o} achieves performance comparable to \text{Claude-4} at lower API cost, and consistently outperforms \text{DS-V3.2}. We therefore adopted \text{GPT-4o}.

\textbf{Threshold Choice.} We set $\epsilon_T=0.01$ seconds and $\epsilon_M=1.5$ MB following prior work~\cite{liu2024evaluating}, using the average resource consumption of a simple \textit{Hello World} program across C++, Java, and Python as the baseline. We set $\epsilon_D=0.05$, which is higher than both the fluctuation range reported in prior work~\cite{peng2025coffe} and the variance observed in our experiment environment (Appendix~\ref{subsection:experiment_environment}). 


\begin{table}[ht]
  \centering
  \footnotesize
  \caption{The average ET (seconds) and PM (MB) of the obtained stress tests across models and iterations.}
  \label{table:pipeline_ablation_model_iteration}
  \setlength{\tabcolsep}{3.5pt}
  \resizebox{\columnwidth}{!}{%
  \begin{tabular}{llcccccc}
    \toprule
    \textbf{LLM} & \textbf{Metric} & \textbf{Iter1} & \textbf{Iter2} & \textbf{Iter3} & \textbf{Iter4} & \textbf{Iter5} & \textbf{Iter6} \\
    \midrule
    \multirow{2}{*}{Claude-4}
    & ET & 0.72 & 1.85 & 2.93 & 3.60 & 4.20 & 4.47 \\
    & PM & 72.3 & 299.5 & 504.6 & 655.2 & 782.9 & 846.3 \\
    \midrule
    \multirow{2}{*}{GPT-4o}
    & ET & 0.69 & 1.79 & 2.78 & 3.57 & 4.09 & 4.37 \\
    & PM & 78.1 & 305.1 & 503.5 & 648.2 & 772.4 & 804.6 \\
    \midrule
    \multirow{2}{*}{DS-V3.2}
    & ET & 0.63 & 1.59 & 2.38 & 3.02 & 3.51 & 3.74 \\
    & PM & 69.2 & 256.2 & 426.2 & 552.0 & 661.8 & 718.3 \\
    \bottomrule
  \end{tabular}%
  }
\end{table}

\subsection{Details of \textsc{Trace}'s Stress Tests}

Table~\ref{table:stress_test_profile_per_iteration} presents the average ET and PM for the top-10 obtained stress tests across iterations. Both efficiency indicators consistently increase with each iteration, confirming that our iterative efficiency-oriented approach successfully guides the LLM to generate more resource-intensive tests. 


\begin{table}[ht]
\centering
\footnotesize
\setlength{\tabcolsep}{2.5pt} 
\caption{The average ET (seconds) and PM (MB) of the top-10 stress tests across iterations.}
\label{table:stress_test_profile_per_iteration}
\begin{tabular}{@{}llrrrrr@{}}
\toprule
\multicolumn{2}{l}{\textbf{Language}} & \textbf{Iter 1} & \textbf{Iter 2} & \textbf{Iter 3} & \textbf{Iter 4} & \textbf{Iter 5} \\
\midrule
\multirow{2}{*}{C++} & ET& 0.14 & 0.24 & 0.33 & 0.42 & \textbf{0.50} \\
& PM & 19.01 & 22.43 & 25.85 & 28.70 & \textbf{31.66} \\
\midrule
\multirow{2}{*}{Java} & ET& 0.50 & 0.69 & 0.88 & 1.08 & \textbf{1.27} \\
& PM & 131.10 & 173.69 & 216.28 & 252.24 & \textbf{279.41} \\
\midrule
\multirow{2}{*}{Python} & ET& 1.45 & 3.93 & 6.41 & 8.29 & \textbf{9.88} \\
& PM & 54.13 & 585.00 & 1115.86 & 1496.83 & \textbf{1897.97} \\
\midrule
\multirow{2}{*}{Avg.} & ET& 0.68 & 1.59 & 2.50 & 3.20 & \textbf{3.80} \\
& PM & 67.36 & 256.95 & 446.55 & 582.27 & \textbf{721.09} \\
\bottomrule
\end{tabular}
\end{table}

Table~\ref{table:test_profile_comparison} compares the effectiveness between \textsc{Transcoder-Test}'s default tests and \textsc{Trace}'s stress tests across LLM-generated reference translations. As shown, \textsc{Trace}'s stress tests significantly amplify the computational demands. On \textsc{Transcoder-Test}'s original tasks, stress tests increase the average ET by a factor of {84.57$\times$} (from 0.07 s to 5.92 s) and PM by {78.6$\times$} (from 15.02 MB to 1180.64 MB). On \textsc{Trace}'s selected tasks, the stress tests still demonstrate a significant superiority, increasing the ET by {8.9$\times$} (from 0.09 s to 0.80 s) and PM by {3.4$\times$} (from 24.35 MB to 83.09 MB). This substantial increase in resource consumption is critical for exposing latent efficiency differences that are otherwise undetectable with the default tests from the original benchmark.

\begin{table}[ht]
\centering
\small
\caption{The performance comparison between Transcoder-Test's default tests and \textsc{Trace}'s stress tests.}
\label{table:test_profile_comparison}
\begin{tabular}{@{}ll rr rr@{}}
\toprule
\multicolumn{2}{c}{\multirow{2}{*}{\textbf{Language}}} & \multicolumn{2}{c}{\textbf{Transcoder-Test}} & \multicolumn{2}{c}{\textbf{TRACE}} \\
\cmidrule(lr){3-4} \cmidrule(lr){5-6}
\multicolumn{2}{c}{} & \textbf{Default} & \textbf{Stress} & \textbf{Default} & \textbf{Stress} \\
\midrule
\multirow{2}{*}{C++} & ET & <0.01 & 0.65 & <0.01 & 0.28 \\
& PM &1.51 & 32.50 & 1.52 & 32.04 \\
\midrule
\multirow{2}{*}{Java} & ET & 0.18 & 2.16 & 0.18 & 0.54 \\
& PM &45.86 & 497.04 & 46.09 & 140.90 \\
\midrule
\multirow{2}{*}{Python} & ET & 0.02 & 14.85 & 0.01 & 1.57 \\
& PM &7.72 & 3030.54 & 5.56 & 25.62 \\
\midrule
\multirow{2}{*}{Avg.} & ET & 0.07 & \textbf{5.92} & 0.09 & \textbf{0.80} \\
& PM &15.02 & \textbf{1180.64} & 24.35 & \textbf{83.09} \\
\bottomrule
\end{tabular}%
\end{table}

\subsection{Details of Task Pruning}
Table~\ref{table:prunned_task_detail} summarizes the task pruning process used to construct \textsc{Trace}. Starting from the original 2,828 translation tasks in \textsc{Transcoder-Test}, we first discard 362 tasks for which no valid stress test suite could be generated. We then apply three filtering rules introduced in the main paper: Rule-1 (Feasibility) removes 217 tasks unsolved by any model; Rule-2 (Impactfulness) excludes 309 tasks where efficiency is not practically relevant; and Rule-3 (Diversity) eliminates 940 tasks where correct solutions show negligible performance variation. After these stages, the benchmark retains {1,000} tasks that are demonstrably efficiency-critical.

\begin{table}[ht]
  \centering
  \footnotesize
  \caption{Statistics of the filtered tasks by each rule.}
  \label{table:prunned_task_detail}
  \setlength{\tabcolsep}{4.0pt}
  \resizebox{\columnwidth}{!}{%
  \begin{tabular}{lrrrrrr}
    \toprule
    \textbf{Direction} & \textbf{Orig.} & \textbf{w/o $\mathcal{T}_s$} & \textbf{Rule-1} & \textbf{Rule-2} & \textbf{Rule-3} & \textbf{\textsc{Trace}} \\
    \midrule
    C++$\rightarrow$Java   & 482 & 59 & 14  & 0   & 159 & 250 \\
    C++$\rightarrow$Py    & 464 & 63 & 32  & 0   & 228 & 141 \\
    Java$\rightarrow$C++   & 468 & 59 & 20  & 160 & 102 & 127 \\
    Java$\rightarrow$Py    & 464 & 63 & 29  & 0   & 224 & 148 \\
    Py$\rightarrow$C++     & 468 & 59 & 87  & 149 & 75  & 98 \\
    Py$\rightarrow$Java    & 482 & 59 & 35  & 0   & 152 & 236 \\
    \midrule
    {Summary} & {2828} & {-362} & {-217} & {-309} & {-940} & {1000} \\
    \bottomrule
  \end{tabular}%
  }
\end{table}

\subsection{Experiment Environment Setup}
\label{subsection:experiment_environment}
All the experiments were conducted in a controlled environment, as summarized in Table~\ref{table:experiment_environment} and Table~\ref{table:experiment_settings}, to ensure consistency. To minimize interference, each profiled command was pinned to the least-utilized CPU core using \texttt{taskset -c}. ET was measured with \texttt{perf stat}, and PM was monitored by sampling the Resident Set Size (RSS) of the process tree via \texttt{psutil}.

We enforced a 180-second timeout and a 4096-MB peak memory limit, calibrated from reference translations to cover diverse valid implementations while excluding pathological cases such as infinite loops. Each execution profile was averaged over five independent runs to mitigate noise from system load. Under this controlled setup, measurement variance was minimal, with coefficients of variation for both ET and PM typically below 3.0\%. Prior work~\cite{peng2025coffe} reported a broader fluctuation range of 2–5\% in less controlled environments. To conservatively filter out noise, we set the diversity threshold to $\epsilon_D=0.05$, which exceeds both our observed variance and the reported noise floor, thereby ensuring that retained tasks reflect genuine algorithmic or idiomatic efficiency differences rather than measurement errors.

\begin{table}[t]
\centering
\footnotesize
\caption{Experiment environment details.}
\label{table:experiment_environment}
\renewcommand{\arraystretch}{0.95}
\begin{tabularx}{\columnwidth}{@{}lX@{}}
\toprule
\textbf{Component} & \textbf{Specification} \\
\midrule
\multicolumn{2}{l}{\best{Hardware}} \\
Operating System & Ubuntu 22.04.5 LTS \\
CPU Type & Intel(R) Xeon(R) Platinum 8468V @ 3.8 GHz \\
CPU Cores & 192 \\
GPU Model & NVIDIA H200 NVL-140GB \\
GPU Quantity & 2 \\
System Memory (RAM) & 512 GB \\
\midrule
\multicolumn{2}{l}{\best{Software}} \\
Linux Kernel & 6.8.0-60-generic \\
C++ Compiler & g++ (Ubuntu 22.04) \\
Java Development Kit & openjdk 17.0.15 \\
Python Interpreter & Python 3.12 \\
\midrule
\multicolumn{2}{l}{\best{Instruction}} \\
C++ Compilation & \texttt{g++ -O2} \\
Java Compilation & \texttt{javac} \\
Time Profiling & \texttt{perf stat} \\
Memory Profiling & Custom script using \texttt{psutil} \\
\bottomrule
\end{tabularx}
\end{table}

\begin{table}[t]
\centering
\footnotesize
\caption{Experiment settings.}
\label{table:experiment_settings}
\renewcommand{\arraystretch}{0.95}
\begin{tabularx}{\columnwidth}{@{}lX@{}}
\toprule
\textbf{Setting} & \textbf{Value} \\
\midrule
Timeout Limit & 180 seconds \\
Memory Limit & 4096 MB peak RSS \\
Runs per Test & 5 (arithmetic mean reported) \\
\bottomrule
\end{tabularx}
\end{table}

\subsection{Evaluation Model Choice}
Table~\ref{table:model_list} provides details for the 28 LLMs used in our work. These models were utilized for both the curation of the benchmark and the evaluation of the work.

\clearpage
\begin{table*}[tbh]
  \centering
  \small
  \caption{Model details, including aliases used in this paper, names, and public links.}
  \label{table:model_list}
  \begin{tabular}{l l l}
    \toprule
    \textbf{Alias} & \textbf{Name} & \textbf{Public Link} \\
    \midrule
    \multicolumn{3}{c}{\textit{Proprietary Models}} \\
    \midrule
    Claude-4-Think    & Claude-Sonnet-4-Think     & \url{https://www.anthropic.com/news/claude-4} \\
    Claude-4          & Claude-Sonnet-4              & \url{https://www.anthropic.com/news/claude-4} \\
    Claude-3.5        & Claude-Sonnet-3.5            & \url{https://www.anthropic.com/news/claude-3-5-sonnet} \\
    DS-Reasoner       & DeepSeek-Reasoner            & \url{https://www.deepseek.com} \\
    DS-V3.2           & DeepSeek-V3.2                & \url{https://www.deepseek.com} \\
    Gemini-Pro        & Gemini-2.5-Pro               & \url{https://deepmind.google/models/gemini/pro} \\
    Gemini-Flash      & Gemini-2.5-Flash             & \url{https://deepmind.google/models/gemini/flash} \\
    GPT-4o            & GPT-4o                       & \url{https://openai.com} \\
    GPT-4o-Mini       & GPT-4o-Mini                  & \url{https://openai.com} \\
    GPT-3.5-Turbo     & GPT-3.5-Turbo                & \url{https://openai.com} \\
    O3                & OpenAI-O3                    & \url{https://openai.com} \\
    O3-Mini           & OpenAI-O3-Mini               & \url{https://openai.com} \\
    \midrule
    \multicolumn{3}{c}{\textit{Open-Source Models}} \\
    \midrule
    CL-7B             & CodeLlama-7B                 & \url{https://github.com/facebookresearch/codellama} \\
    CL-7B-Inst        & CodeLlama-7B-Instruct        & \url{https://github.com/facebookresearch/codellama} \\
    CL-13B            & CodeLlama-13B                & \url{https://github.com/facebookresearch/codellama} \\
    CL-13B-Inst       & CodeLlama-13B-Instruct       & \url{https://github.com/facebookresearch/codellama} \\
    CL-34B            & CodeLlama-34B                & \url{https://github.com/facebookresearch/codellama} \\
    CL-34B-Inst       & CodeLlama-34B-Instruct       & \url{https://github.com/facebookresearch/codellama} \\
    DSC-6.7B          & deepseek-coder-6.7b          & \url{https://github.com/deepseek-ai/DeepSeek-Coder} \\
    DSC-6.7B-Inst     & deepseek-coder-6.7b-Instruct & \url{https://github.com/deepseek-ai/DeepSeek-Coder} \\
    DSC-33B           & deepseek-coder-33b           & \url{https://github.com/deepseek-ai/DeepSeek-Coder} \\
    DSC-33B-Inst      & deepseek-coder-33b-Instruct  & \url{https://github.com/deepseek-ai/DeepSeek-Coder} \\
    QC-7B             & Qwen2.5-Coder-7B             & \url{https://github.com/QwenLM/Qwen2.5-Coder} \\
    QC-7B-Inst        & Qwen2.5-Coder-7B-Instruct    & \url{https://github.com/QwenLM/Qwen2.5-Coder} \\
    QC-14B            & Qwen2.5-Coder-14B            & \url{https://github.com/QwenLM/Qwen2.5-Coder} \\
    QC-14B-Inst       & Qwen2.5-Coder-14B-Instruct   & \url{https://github.com/QwenLM/Qwen2.5-Coder} \\
    QC-32B            & Qwen2.5-Coder-32B            & \url{https://github.com/QwenLM/Qwen2.5-Coder} \\
    QC-32B-Inst       & Qwen2.5-Coder-32B-Instruct   & \url{https://github.com/QwenLM/Qwen2.5-Coder} \\
    \bottomrule
  \end{tabular}
\end{table*}
\begin{table*}[t]
\centering
\caption{Comparison of \textsc{Trace} with representative code translation benchmarks. Corr. and Stress. denote the average number of correctness and stress tests, respectively.}

\footnotesize
\label{table:benchmark_comparison}
\begin{tabular}{p{2.2cm} p{1.5cm} p{0.75cm} p{3.8cm} p{1.5cm} p{0.75cm} p{0.75cm} p{1.5cm}}
\toprule
\textbf{Benchmark} & \textbf{Source} & \textbf{Prob.} & \textbf{Programming Languages} & \textbf{Granularity} & \textbf{Corr.} & \textbf{Stress.} & \textbf{Evaluation} \\
\midrule
CodeNet & Contests & 200 & C++, C, Go, Java, Py & stmt./Method & 1 & \textbf{\XSolidBrush} & Correctness \\
Avatar & Contests & 250 & Java, Py & stmt./Method & 25 & \textbf{\XSolidBrush} & Correctness \\
Transcoder-Test & Geeks4Geeks & 568 & C++, Java, Py & Method & 10 & \textbf{\XSolidBrush} & Correctness \\
G-TransEval & Multiple & 400 & C++, C\#, Java, Py, Js & Method & 5 & \textbf{\XSolidBrush} & Correctness \\
PolyHumanEval & HumanEval & 164 & C++, C\#, Java, Py (+10 others) & Method & 7 & \textbf{\XSolidBrush} & Correctness \\
ClassEval-T & Multiple & 94 & C++, Java, Py & Class & 34 & \textbf{\XSolidBrush} & Correctness \\
\midrule
\textbf{\textsc{Trace}} & Geeks4Geeks & 357 & C++, Java, Py & Method & 10 & 10 & \begin{tabular}{@{}l@{}}\best{Efficiency} \\ Correctness\end{tabular} \\
\bottomrule
\end{tabular}
\end{table*}
\clearpage
\clearpage

\section{Extended Evaluation}

\subsection{Benchmark Comparison}
Table~\ref{table:benchmark_comparison} compares \textsc{Trace} with representative code translation benchmarks, which have traditionally centered on functional correctness. Benchmarks such as CodeNet~\cite{puri2021codenet}, Avatar~\cite{ahmad2021avatar}, Transcoder-Test~\cite{lachaux2020unsupervised}, G-TransEval~\cite{jiao2023evaluation}, PolyHumanEval~\cite{tao2024unraveling}, and ClassEval-T~\cite{xue2025classeval} provide only correctness-oriented tests and lack mechanisms to capture efficiency differences. Thus, they may fail to reveal runtime efficiency differences that often remain hidden under small-scale evaluation. In contrast, \textsc{Trace} augments each task with stress tests that exercise computational complexity, and selectively retains efficiency-critical tasks to ensure meaningful differentiation. 


\subsection{Granular Performance Comparison}
\label{section:granular_performance_comparison}

To complement the overall results in the main paper, Table~\ref{table:granular_performance_comparison} reports a per-direction breakdown of correctness and efficiency across six translation directions, offering a fine-grained view of model behavior beyond aggregate performance.

\subsection{Statistical Relationship Between Correctness and Efficiency}
\label{section:relationship_correctness_and_efficiency}

To understand whether functional correctness aligns with efficiency, we conduct correlation and regression analysis across all evaluated LLMs. Since incorrect translations trivially receive zero efficiency, we report conditioned efficiency scores ($B_T^P$, $B_M^P$) that only consider successful cases.

Table~\ref{table:correlation_stats} summarizes the correlation results. Across the full set of models, correctness is moderately negatively correlated with time efficiency (Pearson $r=-0.54$, Spearman $\rho=-0.61$) and moderately positively correlated with memory efficiency ($r=0.57$, $\rho=0.70$). Similar trends hold for the high-performing subset (Pass $\geq 85\%$, 613 tasks in common), where correctness again shows a negative association with time efficiency ($r=-0.51$) and a positive association with memory efficiency ($r=0.64$). These findings demonstrate that higher correctness does not necessarily imply higher efficiency.

\begin{table}[htbp]
\centering
\footnotesize
\caption{Correlation analysis between correctness (Pass) and efficiency (Beyond); † indicates $p<0.05$.}
\footnotesize
\label{table:correlation_stats}
\begin{tabular*}{\columnwidth}{@{\extracolsep{\fill}}lcc}
\toprule
\textbf{Correctness v.s. Efficiency} & \textbf{Pearson $r$} & \textbf{Spearman $\rho$} \\
\midrule
\multicolumn{3}{l}{\best{Full Set (28 models, 1000 tasks)}} \\
Pass vs $B_T^P$     & $-0.54^{\dagger}$ & $-0.61^{\dagger}$ \\
Pass vs $B_M^P$     & $0.57^{\dagger}$  & $0.70^{\dagger}$ \\
\midrule
\multicolumn{3}{l}{\best{High-Pass Set (16 models, 613 tasks)}} \\
Pass vs $B_T^{com}$ & $-0.51^{\dagger}$ & $-0.43^{\dagger}$ \\
Pass vs $B_M^{com}$ & $0.64^{\dagger}$  & $0.60^{\dagger}$ \\
\bottomrule
\end{tabular*}
\end{table}

To further quantify explanatory power, we fit Ordinary Least Squares (OLS) regressions with efficiency as the dependent variable and correctness as the predictor. The fitted models yield a negative slope for time efficiency ($\beta_1=-0.153$, $R^2=0.29$) and a positive slope for memory efficiency ($\beta_1=0.236$, $R^2=0.33$). The relatively low $R^2$ values indicate that correctness accounts for only a limited portion of efficiency variance, confirming that efficiency reflects distinct factors beyond functional accuracy and should be evaluated as an independent dimension.

\begin{figure}[htbp]
    \centering
    \includegraphics[width=0.7\linewidth]{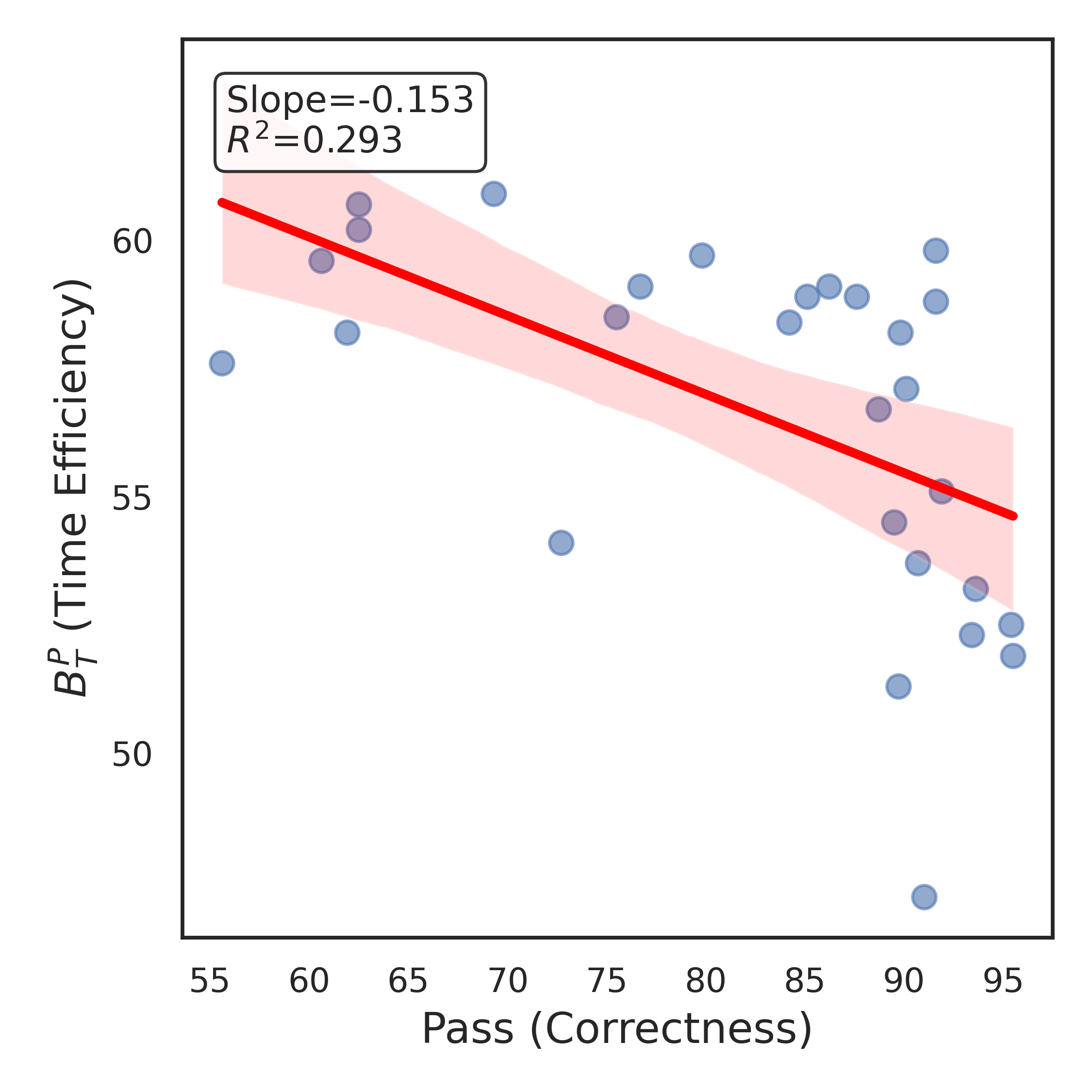} \\
    \includegraphics[width=0.7\linewidth]{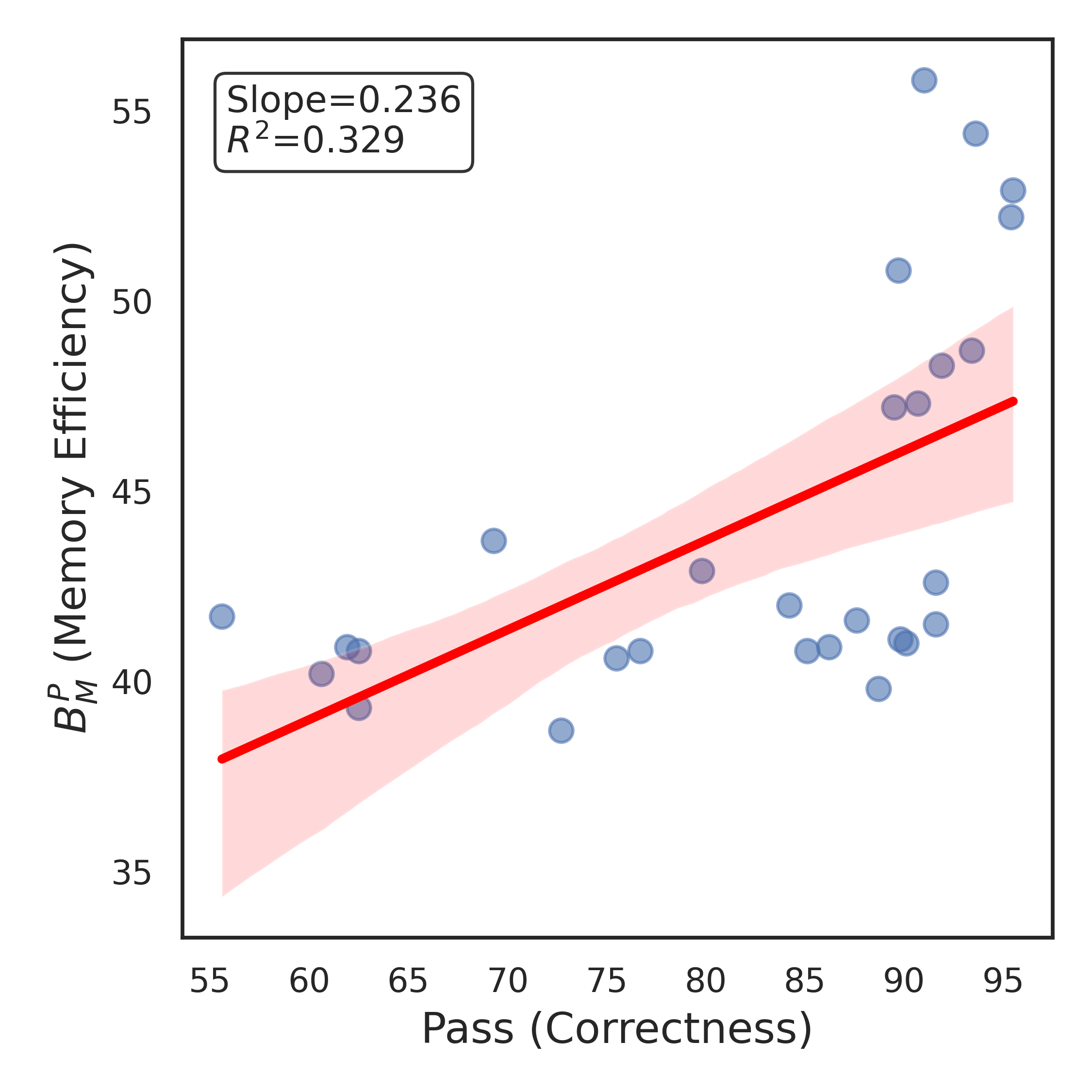}
    \caption{OLS regression of correctness-conditioned Beyond score (efficiency) on Pass rate (correctness).}
    \label{figure:appendix_regression}
\end{figure}

\subsection{Additional Case Study}
\label{subsection:case_study}
To provide concrete illustrations of the inefficiency patterns in our taxonomy, we further present three representative cases. Each case corresponds to one of the high-level categories and demonstrates dramatic performance degradation under stress testing. The first case study (Figure~\ref{figure:case_study_algorithm}) highlights a critical algorithmic degradation. The second (Figure~\ref{figure:case_study_idiom}) shows the impact of suboptimal non-idiomatic data structure. The final case (Figure~\ref{figure:case_study_overhead}) illustrates how unnecessary resource overhead can severely harm performance.

\clearpage

\begin{table*}[ht]
\centering
\small
\caption{Granular LLM efficiency performance across six translation directions. For each direction, we report \textbf{Pass} , \(\boldsymbol{B_T}\), \(\boldsymbol{B_M}\) \(\boldsymbol{B_{T}^P}\), and \(\boldsymbol{B_{M}^P}\) (\%).}
\label{table:granular_performance_comparison}
\setlength{\tabcolsep}{3.8pt}
\renewcommand{\arraystretch}{0.95}

\begin{tabular}{l *{15}{c}}
\toprule
\multirow{2}{*}{\textbf{Model}}
& \multicolumn{5}{c}{\textbf{C++$\rightarrow$Java}}
& \multicolumn{5}{c}{\textbf{C++$\rightarrow$Py}}
& \multicolumn{5}{c}{\textbf{Java$\rightarrow$C++}} \\
\cmidrule(lr){2-6} \cmidrule(lr){7-11} \cmidrule(lr){12-16}
& \textbf{Pass} & $\boldsymbol{B_T}$ & $\boldsymbol{B_M}$ & $\boldsymbol{B_{T}^P}$ & $\boldsymbol{B_{M}^P}$
& \textbf{Pass} & $\boldsymbol{B_T}$ & $\boldsymbol{B_M}$ & $\boldsymbol{B_{T}^P}$ & $\boldsymbol{B_{M}^P}$
& \textbf{Pass} & $\boldsymbol{B_T}$ & $\boldsymbol{B_M}$ & $\boldsymbol{B_{T}^P}$ & $\boldsymbol{B_{M}^P}$ \\
\midrule

Claude-4-Think & 95.6 & 37.1 & 54.0 & 38.8 & 56.5 & 97.2 & 61.3 & 48.4 & 63.1 & 49.8 & 98.4 & 58.8 & 50.7 & 59.8 & 51.6 \\
Claude-4       & 97.2 & 66.8 & 62.9 & 68.8 & 64.7 & 97.2 & 25.7 & 13.2 & 26.5 & 13.6 & 91.3 & 45.5 & 56.6 & 49.9 & 62.0 \\
Claude-3.5     & 98.0 & 45.2 & 51.9 & 46.1 & 52.9 & 97.2 & 64.5 & 50.5 & 66.4 & 51.9 & 92.1 & 53.6 & 46.2 & 58.2 & 50.2 \\
DS-Reasoner    & 81.2 & 34.6 & 28.4 & 42.6 & 35.0 & 78.0 & 45.3 & 36.0 & 58.1 & 46.1 & 61.4 & 31.4 & 23.9 & 51.2 & 39.0 \\
DS-V3.2        & 97.2 & 67.2 & 63.3 & 69.2 & 65.1 & 95.7 & 30.0 & 16.3 & 31.3 & 17.0 & 85.0 & 49.4 & 46.5 & 58.1 & 54.7 \\
Gemini-Pro     & 58.8 & 35.9 & 20.7 & 61.1 & 35.2 & 70.2 & 44.8 & 34.7 & 63.8 & 49.4 & 56.7 & 29.3 & 22.2 & 51.6 & 39.1 \\
Gemini-Flash   & 68.4 & 37.2 & 26.2 & 54.3 & 38.2 & 73.0 & 49.8 & 37.0 & 68.1 & 50.6 & 55.1 & 33.2 & 22.6 & 60.2 & 41.0 \\
GPT-4o         & 95.2 & 63.9 & 63.0 & 67.1 & 66.2 & 97.2 & 31.6 & 9.7  & 32.5 & 10.0 & 66.9 & 36.1 & 35.7 & 53.9 & 53.3 \\
GPT-4o-Mini    & 91.2 & 37.7 & 50.6 & 41.3 & 55.5 & 95.0 & 59.7 & 43.2 & 62.9 & 45.4 & 81.9 & 47.7 & 42.3 & 58.2 & 51.7 \\
O3             & 94.0 & 53.5 & 34.3 & 57.0 & 36.5 & 95.0 & 63.6 & 49.0 & 66.9 & 51.5 & 68.5 & 39.9 & 25.1 & 58.3 & 36.7 \\
O3-Mini        & 95.2 & 40.7 & 52.3 & 42.8 & 55.0 & 95.7 & 37.4 & 63.5 & 39.0 & 66.3 & 89.0 & 48.9 & 47.3 & 54.9 & 53.1 \\
GPT-3.5-Turbo        & 95.6 & 43.6 & 54.6 & 45.6 & 57.2 & 92.2 & 61.0 & 45.5 & 66.2 & 49.3 & 91.3 & 56.5 & 52.1 & 61.9 & 57.0 \\
\midrule
CL-7B          & 57.2 & 33.8 & 22.0 & 59.0 & 38.5 & 67.4 & 44.2 & 30.9 & 65.7 & 45.8 & 47.2 & 24.9 & 19.9 & 52.7 & 42.0 \\
CL-7B-Inst     & 75.2 & 39.1 & 28.6 & 51.9 & 38.0 & 82.3 & 53.5 & 39.2 & 65.1 & 47.6 & 70.9 & 43.1 & 33.2 & 60.8 & 46.9 \\
CL-13B         & 62.8 & 33.7 & 22.1 & 53.7 & 35.2 & 68.1 & 43.0 & 32.9 & 63.1 & 48.4 & 36.2 & 19.8 & 18.4 & 54.8 & 50.9 \\
CL-13B-Inst    & 73.2 & 40.7 & 26.0 & 55.7 & 35.6 & 80.9 & 53.0 & 39.8 & 65.6 & 49.3 & 77.2 & 43.7 & 34.9 & 56.7 & 45.2 \\
CL-34B         & 66.4 & 34.9 & 25.4 & 52.6 & 38.3 & 62.4 & 42.6 & 30.7 & 68.3 & 49.2 & 14.2 & 9.5  & 6.8  & 67.2 & 47.9 \\
CL-34B-Inst    & 65.6 & 36.6 & 24.1 & 55.8 & 36.8 & 83.0 & 54.8 & 43.4 & 66.0 & 52.2 & 50.4 & 30.4 & 29.2 & 60.4 & 58.0 \\
DSC-6.7B       & 76.8 & 40.8 & 30.1 & 53.1 & 39.2 & 87.9 & 56.6 & 41.3 & 64.4 & 47.0 & 75.6 & 45.2 & 33.2 & 59.8 & 43.9 \\
DSC-6.7B-Inst  & 90.4 & 50.9 & 33.7 & 56.3 & 37.3 & 89.4 & 58.4 & 44.9 & 65.4 & 50.2 & 74.0 & 42.4 & 34.2 & 57.3 & 46.1 \\
DSC-33B        & 88.8 & 47.9 & 37.1 & 53.9 & 41.8 & 87.2 & 56.8 & 42.3 & 65.1 & 48.5 & 70.1 & 37.1 & 28.2 & 53.0 & 40.2 \\
DSC-33B-Inst   & 95.2 & 54.8 & 36.5 & 57.5 & 38.3 & 93.6 & 60.6 & 46.9 & 64.7 & 50.1 & 76.4 & 41.3 & 27.2 & 54.0 & 35.6 \\
QC-7B          & 89.6 & 48.1 & 31.9 & 53.7 & 35.6 & 92.9 & 58.3 & 46.0 & 62.8 & 49.5 & 84.3 & 48.0 & 36.5 & 57.0 & 43.3 \\
QC-7B-Inst     & 94.0 & 42.5 & 33.0 & 45.2 & 35.1 & 92.2 & 64.3 & 44.8 & 69.7 & 48.6 & 85.8 & 48.5 & 39.2 & 56.6 & 45.7 \\
QC-14B         & 91.6 & 52.0 & 33.3 & 56.7 & 36.3 & 94.3 & 64.4 & 46.4 & 68.2 & 49.1 & 88.2 & 51.6 & 45.6 & 58.5 & 51.7 \\
QC-14B-Inst    & 96.0 & 54.3 & 36.5 & 56.5 & 38.1 & 97.2 & 63.3 & 46.5 & 65.1 & 47.9 & 81.9 & 46.2 & 37.4 & 56.5 & 45.7 \\
QC-32B         & 92.0 & 52.1 & 34.9 & 56.6 & 37.9 & 91.5 & 59.5 & 43.9 & 65.0 & 48.0 & 84.3 & 46.3 & 33.7 & 54.9 & 40.0 \\
QC-32B-Inst    & 95.2 & 56.2 & 37.7 & 59.0 & 39.6 & 93.6 & 63.8 & 44.7 & 68.1 & 47.7 & 83.5 & 50.3 & 40.6 & 60.3 & 48.7 \\
\midrule
Average        & 84.9 & 45.8 & 37.7 & 54.0 & 43.6 & 87.4 & 52.6 & 39.7 & 60.6 & 45.7 & 72.8 & 41.4 & 34.6 & 57.0 & 47.2 \\
\bottomrule
\end{tabular}

\vspace{1em}

\begin{tabular}{l *{15}{c}}
\toprule
\multirow{2}{*}{\textbf{Model}}
& \multicolumn{5}{c}{\textbf{Java$\rightarrow$Py}}
& \multicolumn{5}{c}{\textbf{Py$\rightarrow$C++}}
& \multicolumn{5}{c}{\textbf{Py$\rightarrow$Java}} \\
\cmidrule(lr){2-6} \cmidrule(lr){7-11} \cmidrule(lr){12-16}
& \textbf{Pass} & $\boldsymbol{B_T}$ & $\boldsymbol{B_M}$ & $\boldsymbol{B_{T}^P}$ & $\boldsymbol{B_{M}^P}$
& \textbf{Pass} & $\boldsymbol{B_T}$ & $\boldsymbol{B_M}$ & $\boldsymbol{B_{T}^P}$ & $\boldsymbol{B_{M}^P}$
& \textbf{Pass} & $\boldsymbol{B_T}$ & $\boldsymbol{B_M}$ & $\boldsymbol{B_{T}^P}$ & $\boldsymbol{B_{M}^P}$ \\
\midrule
Claude-4-Think & 98.6 & 65.4 & 46.9 & 66.3 & 47.6 & 86.7 & 52.0 & 44.4 & 59.9 & 51.2 & 94.5 & 39.9 & 52.6 & 42.2 & 55.6 \\
Claude-4       & 98.0 & 27.0 & 16.1 & 27.6 & 16.4 & 81.6 & 43.4 & 47.5 & 53.2 & 58.1 & 90.3 & 61.1 & 58.1 & 67.7 & 64.3 \\
Claude-3.5     & 96.6 & 61.2 & 45.0 & 63.3 & 46.6 & 85.7 & 46.1 & 44.1 & 53.8 & 51.4 & 96.6 & 39.5 & 54.6 & 40.9 & 56.6 \\
DS-Reasoner    & 80.4 & 53.8 & 36.9 & 67.0 & 45.9 & 57.1 & 32.8 & 19.3 & 57.5 & 33.7 & 68.2 & 38.5 & 23.5 & 56.4 & 34.4 \\
DS-V3.2        & 96.6 & 33.0 & 14.9 & 34.2 & 15.5 & 66.3 & 30.4 & 31.7 & 45.8 & 47.7 & 95.3 & 65.6 & 63.7 & 68.8 & 66.8 \\
Gemini-Pro     & 73.0 & 47.6 & 35.8 & 65.3 & 49.0 & 42.9 & 27.3 & 10.0 & 63.8 & 23.3 & 66.5 & 38.9 & 22.8 & 58.4 & 34.3 \\
Gemini-Flash   & 79.1 & 51.3 & 35.0 & 64.8 & 44.2 & 29.6 & 18.1 & 7.9  & 61.0 & 26.5 & 57.2 & 32.8 & 20.9 & 57.3 & 36.5 \\
GPT-4o         & 97.3 & 31.0 & 15.4 & 31.9 & 15.9 & 64.3 & 35.2 & 32.5 & 54.8 & 50.6 & 96.6 & 66.7 & 64.0 & 69.1 & 66.3 \\
GPT-4o-Mini    & 97.3 & 63.7 & 41.7 & 65.5 & 42.9 & 66.3 & 35.4 & 30.8 & 53.3 & 46.5 & 94.1 & 39.2 & 51.8 & 41.7 & 55.1 \\
O3             & 93.9 & 61.0 & 46.5 & 65.0 & 49.5 & 52.0 & 26.9 & 17.2 & 51.7 & 33.0 & 86.9 & 46.9 & 31.8 & 54.0 & 36.6 \\
O3-Mini        & 95.3 & 58.4 & 46.1 & 61.3 & 48.4 & 62.2 & 33.0 & 32.2 & 53.0 & 51.7 & 94.1 & 39.7 & 54.0 & 42.2 & 57.4 \\
GPT-3.5-Turbo        & 99.3 & 58.8 & 45.7 & 59.2 & 46.0 & 83.7 & 47.4 & 49.5 & 56.7 & 59.2 & 94.1 & 41.4 & 53.5 & 44.0 & 56.8 \\
\midrule

CL-7B          & 70.9 & 46.7 & 32.4 & 65.8 & 45.6 & 54.1 & 32.1 & 22.1 & 59.4 & 40.8 & 69.1 & 34.6 & 25.3 & 50.1 & 36.6 \\
CL-7B-Inst     & 81.8 & 53.7 & 38.0 & 65.7 & 46.5 & 53.1 & 34.2 & 15.8 & 64.4 & 29.8 & 79.7 & 42.7 & 27.7 & 53.6 & 34.8 \\
CL-13B         & 70.3 & 46.6 & 31.8 & 66.4 & 45.2 & 33.7 & 20.5 & 11.5 & 61.0 & 34.3 & 72.0 & 43.1 & 25.5 & 59.8 & 35.5 \\
CL-13B-Inst    & 84.5 & 56.5 & 39.2 & 66.9 & 46.4 & 58.2 & 32.8 & 26.5 & 56.4 & 45.5 & 80.5 & 44.7 & 27.0 & 55.5 & 33.5 \\
CL-34B         & 75.7 & 50.7 & 33.5 & 67.0 & 44.3 & 19.4 & 13.9 & 9.6  & 71.6 & 49.5 & 64.8 & 30.6 & 24.4 & 47.2 & 37.7 \\
CL-34B-Inst    & 83.1 & 56.3 & 40.1 & 67.7 & 48.3 & 45.9 & 30.0 & 20.6 & 65.3 & 44.8 & 76.3 & 43.3 & 27.3 & 56.8 & 35.8 \\
DSC-6.7B       & 85.8 & 58.7 & 40.1 & 68.4 & 46.7 & 74.5 & 50.0 & 37.7 & 67.1 & 50.6 & 78.8 & 42.8 & 29.6 & 54.2 & 37.6 \\
DSC-6.7B-Inst  & 93.9 & 62.1 & 43.6 & 66.1 & 46.4 & 60.2 & 36.4 & 22.5 & 60.5 & 37.5 & 92.4 & 50.3 & 31.9 & 54.5 & 34.5 \\
DSC-33B        & 86.5 & 58.2 & 39.0 & 67.3 & 45.1 & 75.5 & 48.6 & 30.5 & 64.3 & 40.3 & 87.3 & 47.0 & 32.9 & 53.8 & 37.7 \\
DSC-33B-Inst   & 96.6 & 64.3 & 46.1 & 66.6 & 47.7 & 73.5 & 40.1 & 30.5 & 54.6 & 41.6 & 91.5 & 47.9 & 33.6 & 52.3 & 36.7 \\
QC-7B          & 93.9 & 63.5 & 40.2 & 67.7 & 42.8 & 68.4 & 38.4 & 26.1 & 56.2 & 38.2 & 92.8 & 45.8 & 32.5 & 49.3 & 35.0 \\
QC-7B-Inst     & 92.6 & 64.9 & 47.4 & 70.1 & 51.2 & 74.5 & 43.2 & 28.3 & 58.0 & 37.9 & 91.9 & 49.6 & 32.1 & 54.0 & 34.9 \\
QC-14B         & 95.9 & 63.9 & 44.9 & 66.6 & 46.7 & 77.6 & 44.6 & 30.6 & 57.6 & 39.4 & 94.9 & 48.2 & 32.8 & 50.8 & 34.6 \\
QC-14B-Inst    & 97.3 & 65.8 & 46.0 & 67.6 & 47.3 & 70.4 & 42.4 & 29.4 & 60.2 & 41.7 & 94.1 & 53.0 & 37.7 & 56.4 & 40.1 \\
QC-32B         & 89.9 & 59.3 & 41.6 & 66.0 & 46.3 & 63.3 & 37.2 & 30.0 & 58.8 & 47.4 & 91.1 & 50.4 & 34.5 & 55.4 & 37.8 \\
QC-32B-Inst    & 96.6 & 33.8 & 68.2 & 35.0 & 70.6 & 64.3 & 26.7 & 31.7 & 41.5 & 49.3 & 95.3 & 49.3 & 37.3 & 51.7 & 39.1 \\
\midrule
Average        & 89.3 & 54.2 & 39.2 & 61.2 & 44.1 & 62.3 & 35.7 & 27.5 & 57.9 & 42.9 & 85.2 & 45.5 & 37.3 & 53.5 & 43.0 \\
\bottomrule
\end{tabular}
\end{table*}

\clearpage

\begin{figure*}[t]
  \centering

  \begin{minipage}[t]{\linewidth}
    \begin{bluebox}[title=\textbf{Java Source Code}]
\begin{minted}{java}
static int findS ( int s ) {
  int sum = 0 ;
  for ( int n = 1 ; sum < s ; n ++ ) {
    sum += n * n ;
    if ( sum == s ) 
        return n ;
  }
  return - 1 ;
}
\end{minted}
    \end{bluebox}
  \end{minipage}

  \vspace{0.8em} %
  \begin{minipage}[t]{\linewidth}
    \begin{bluebox}[title=\textbf{Efficient Python Translation (Claude-Sonnet-4)}]
\begin{minted}{py}
# Execution Time  0.02 s 
# Peak Memory     9.00 MB
def findS(s):
    sum = 0
    n = 1
    while sum < s:
        sum += n * n
        if sum == s:
            return n
        n += 1
    return -1
\end{minted}
    \end{bluebox}
  \end{minipage}

  \vspace{0.8em}

  \begin{minipage}[t]{\linewidth}
    \begin{bluebox}[title=\textbf{Inefficient Python Translation (CodeLlama-13B-Instruct-hf)}]
\begin{minted}{py}
# Execution Time  115.86 s 
# Peak Memory       9.00 MB
def findS(s):
    sum = 0
    for n in range(1, s + 1): # Algorithm Discrepancy
        sum += n * n
        if sum == s:
            return n
    return -1
\end{minted}
    \end{bluebox}
  \end{minipage}
  \caption{A case study of \text{Algorithm Implementation Discrepancy}. The inefficient translation mistakenly converts an early-exit loop into one that iterates over the entire potential range, which degrades the original algorithm's time complexity and causes a catastrophic time slowdown of over \text{6400$\times$}.}
  \label{figure:case_study_algorithm}
\end{figure*}
\begin{figure*}[t]
  \centering

  \begin{minipage}[t]{\linewidth}
    \begin{bluebox}[title=\textbf{Python Source Code}]
\begin{minted}{python}
def findSubarraySum ( arr , n ) :
    res = 0
    m = dict ( )
    for i in range ( n ) :
        Sum = 0
        for j in range ( i , n ) :
            Sum += arr [ j ]
            m [ Sum ] = m.get ( Sum , 0 ) + 1
    for x in m :
        if m [ x ] == 1 :
            res += x
    return res
\end{minted}
    \end{bluebox}
  \end{minipage}

  \vspace{0.8em} %
  \begin{minipage}[t]{\linewidth}
    \begin{bluebox}[title=\textbf{Efficient C++ Translation (Qwen2.5-Coder-7B-Instruct)}]
\begin{minted}{cpp}
// Execution Time  2.09 s 
// Peak Memory    14.60 MB
int findSubarraySum(int arr[], int n) {
    int res = 0;
    unordered_map<int, int> m;
    for (int i = 0; i < n; i++) {
        int Sum = 0;
        for (int j = i; j < n; j++) {
            Sum += arr[j];
            m[Sum]++;
        }
    }
    for (auto x : m) {
        if (x.second == 1) {
            res += x.first;
        }
    }
    return res;
}
\end{minted}
    \end{bluebox}
  \end{minipage}

  \vspace{0.8em}

  \begin{minipage}[t]{\linewidth}
    \begin{bluebox}[title=\textbf{Inefficient C++ Translation (CodeLlama-34B-hf)}]
\begin{minted}{cpp}
// Execution Time  14.40 s 
// Peak Memory     16.50 MB
int findSubarraySum ( int arr [ ] , int n ) {
    int res = 0;
    map < int , int > m; // Language-Construct Misalignment
    for ( int i = 0 ; i < n ; i ++ ) {
        int Sum = 0;
        for ( int j = i ; j < n ; j ++ ) {
            Sum += arr [ j ];
            m [ Sum ] ++;
        }
    }
    for ( auto x : m ) {
        if ( x.second == 1 ) {
            res += x.first;
        }
    }
    return res;
}
\end{minted}
    \end{bluebox}
  \end{minipage}
    \caption{A case study of Language Construct Mismatch. The inefficient translation fails to adopt the hash-based map in C++, relying on the sub-optimal tree-based map instead. This idiomatic misalignment degrades the expected constant-time lookups to logarithmic-time operations, causing a significant slowdown of over \text{6.9$\times$}.}

  \label{figure:case_study_idiom}
\end{figure*}
\begin{figure*}[t]
  \centering

  \begin{minipage}[t]{\linewidth}
    \begin{bluebox}[title=\textbf{C++ Source Code}]
\begin{minted}{cpp}
bool isPrime ( int p ) {
  long long checkNumber = pow ( 2, p ) - 1;
  long long nextval = 4 % checkNumber;
  for ( int i = 1; i < p - 1; i ++ ) 
    nextval = ( nextval * nextval - 2 ) % checkNumber;
  return ( nextval == 0 );
}
\end{minted}
    \end{bluebox}
  \end{minipage}

  \vspace{0.8em} %
  \begin{minipage}[t]{\linewidth}
    \begin{bluebox}[title=\textbf{Efficient Java Translation (Deepseek-V3.2)}]
\begin{minted}{java}
// Execution Time  0.16 s 
// Peak Memory    46.50 MB
public static boolean isPrime(int p) {
    long checkNumber = (long) (Math.pow(2, p) - 1);
    long nextval = 4 % checkNumber;
    for (int i = 1; i < p - 1; i++) {
        nextval = (nextval * nextval - 2) % checkNumber;
    }
    return (nextval == 0);
}
\end{minted}
    \end{bluebox}
  \end{minipage}

  \vspace{0.8em}

  \begin{minipage}[t]{\linewidth}
    \begin{bluebox}[title=\textbf{Inefficient Java Translation (Deepseek-Coder-6.7B-Instruct)}]
\begin{minted}{java}
// Execution Time  8.35 s 
// Peak Memory  1403.50 MB
public static boolean isPrime(int p) {
    BigInteger checkNumber = BigInteger.valueOf(2).pow(p).subtract(BigInteger.ONE); // Resource-Overhead Inefficiency
    BigInteger nextval = BigInteger.valueOf(4).mod(checkNumber);
    for (int i = 1; i < p - 1; i ++) {
        nextval = nextval.multiply(nextval).subtract(BigInteger.valueOf(2)).mod(checkNumber);
    }
    return nextval.compareTo(BigInteger.ZERO) == 0;
}
\end{minted}
    \end{bluebox}
  \end{minipage}

    \caption{A case study of Resource Management Inefficiency. The inefficient translation replaces primitive \texttt{long}-based arithmetic with Java's heavyweight \texttt{BigInteger} class. While functionally correct, this introduces substantial object creation and garbage-collection overhead, resulting in a severe slowdown of over \text{53.9$\times$} and a memory consumption increase of over \text{30.2$\times$}.}

  \label{figure:case_study_overhead}
\end{figure*}

\clearpage

\clearpage

\section{Prompt and Script Design}
\label{section:prompt_and_script_design}

\subsection{Prompt for LLM-based Code Translation}
\label{section:translation_prompt}

We present the prompt templates used in our study. Four settings are considered: 
1) \textit{zero-shot}, which directly asks the LLM to translate code without additional guidance (Figure~\ref{figure:prompt_code_translation_zeroshot}); 
2) \textit{perf-zero-shot}, which augments the zero-shot setting with explicit efficiency-oriented system instructions (Figure~\ref{figure:prompt_code_translation_zeroshoteffi}); 
3) \textit{perf-few-shot}, which provides the LLM with two efficient code translation examples, which are drawn from other problems in \textsc{Trace} following the same translation direction (Figure~\ref{figure:prompt_code_translation_fewshoteffi}); and 
4) \textit{perf-self-refine}, which includes the execution feedback of the translation produced in the previous zero-shot round and asks the LLM for further refinement or optimization (Figure~\ref{figure:prompt_code_translation_selfrefineffi}).

\begin{figure*}[t]
    \centering
    \begin{minipage}{1.0\textwidth}
        \begin{pinkbox}{System Prompt}
\begin{minted}{markdown}
You are a code translation expert proficient in C++, Java, and Python. You translate code accurately while preserving its functionality.
\end{minted}
        \end{pinkbox}
    \end{minipage}
    \vspace{1.0em}
    \begin{minipage}{1.0\textwidth}
        \begin{pinkbox}{User Prompt}
\begin{minted}{markdown}
### Task Description
Given the SRC_LANG code:
```SRC_LANG
SRC_CODE_FILL
```
Please translate the above SRC_LANG code to TGT_LANG code.

### Output Format
Wrap your output in a Markdown code block using ``` backticks, and end your response with <|END|>, e.g.:
```TGT_LANG
...
```
<|END|>
\end{minted}
        \end{pinkbox}
    \end{minipage}
    \caption{The prompt template for the zero-shot setting.}
    \label{figure:prompt_code_translation_zeroshot}
\end{figure*}

\begin{figure*}[t]
    \centering
    \begin{minipage}{1.0\textwidth}
        \begin{pinkbox}{System Prompt}
\begin{minted}{markdown}
You are a code translation expert proficient in C++, Java, and Python. You translate code accurately while preserving its functionality.
You are also a code optimization expert, ensuring the translated code is efficient and adheres to best practices.
\end{minted}
        \end{pinkbox}
    \end{minipage}
    \vspace{1.0em}
    \begin{minipage}{1.0\textwidth}
        \begin{pinkbox}{User Prompt}
\begin{minted}{markdown}
### Task Description
Given the SRC_LANG code:
```SRC_LANG
SRC_CODE_FILL
```
Please translate the above SRC_LANG code to TGT_LANG code.

### Output Format
Wrap your output in a Markdown code block using ``` backticks, and end your response with <|END|>, e.g.:
```TGT_LANG
...
```
<|END|>
\end{minted}
        \end{pinkbox}
    \end{minipage}
    \caption{The prompt template for the perf-zero-shot setting.}
    \label{figure:prompt_code_translation_zeroshoteffi}
\end{figure*}

\begin{figure*}[t]
    \centering
    \begin{minipage}{1.0\textwidth}
        \begin{pinkbox}{System Prompt}
\begin{minted}{markdown}
You are a code translation expert proficient in C++, Java, and Python. You translate code accurately while preserving its functionality.
You are also a code optimization expert, ensuring the translated code is efficient and adheres to best practices.
\end{minted}
        \end{pinkbox}
    \end{minipage}
    \vspace{1.0em}
    \begin{minipage}{1.0\textwidth}
        \begin{pinkbox}{User Prompt}
\begin{minted}{markdown}
### Task Description
Given the SRC_LANG code:
```SRC_LANG
SRC_CODE_FILL
```
Please translate the above SRC_LANG code to TGT_LANG code.

### Translation Examples
Below are some examples of efficient code translations from SRC_LANG to TGT_LANG.

#### Example-1
Given the SRC_LANG code:
```SRC_LANG
SRC_CODE_FILL_EXAMPLE_1
```
The correct and efficient TGT_LANG code translation is as below:
```TGT_LANG
EFFI_TGT_CODE_FILL_1
```
<|END|>

#### Example-2
Given the SRC_LANG code:
```SRC_LANG
SRC_CODE_FILL_EXAMPLE_2
```
The correct and efficient TGT_LANG code translation is as below:
```TGT_LANG
EFFI_TGT_CODE_FILL_2
```
<|END|>

### Output Format
Wrap your output in a Markdown code block using ``` backticks, and end your response with <|END|>, e.g.:
```TGT_LANG
...
```
<|END|>
\end{minted}
        \end{pinkbox}
    \end{minipage}
    \caption{The prompt template for the perf-few-shot setting.}
    \label{figure:prompt_code_translation_fewshoteffi}
\end{figure*}

\begin{figure*}[t]
    \centering
    \begin{minipage}{1.0\textwidth}
        \begin{pinkbox}{System Prompt}
\begin{minted}{markdown}
You are a code translation expert proficient in C++, Java, and Python. You translate code accurately while preserving its functionality.
You are also a code optimization expert, ensuring the translated code is efficient and adheres to best practices.
\end{minted}
        \end{pinkbox}
    \end{minipage}
    \vspace{1.0em}
    \begin{minipage}{1.0\textwidth}
        \begin{pinkbox}{User Prompt}
\begin{minted}{markdown}
### Task
Given the SRC_LANG code:
```SRC_LANG
SRC_CODE_FILL
```
Please translate the above SRC_LANG code to TGT_LANG code. 

### Reference Translation
Below is a reference TGT_LANG translation for the given SRC_LANG code:
```TGT_LANG
REF_TGT_CODE_FILL
```
<|END|>

The execution profile of the reference TGT_LANG translation is as below:
```markdown
**Test Result** (0.0 means all tests failed, and 1.0 means all tests passed)
TEST_RESULT_FILL

**Execution Feedback**
EXEC_FEEDBACK_FILL

**Execution Time (seconds)**
EXEC_TIME_FILL

**Execution Memory (MB)**
EXEC_MEM_FILL
```

If you think the reference TGT_LANG translation is not correct or efficient, please try to improve it. Otherwise, output the original reference TGT_LANG translation without any changes.

### Output Format
Wrap your output in a Markdown code block using ``` backticks, and end your response with <|END|>, e.g.:
```TGT_LANG
...
```
<|END|>
\end{minted}
        \end{pinkbox}
    \end{minipage}
    \caption{The prompt template for the perf-self-refine setting.}
    \label{figure:prompt_code_translation_selfrefineffi}
\end{figure*}

\subsection{Prompt for Stress Test Generation}
Figure~\ref{figure:prompt_stress_test_generation} presents the prompt template used to guide the LLM in generating stress test input synthesizers. The prompt is structured with five key sections: it defines the \texttt{Task} by providing the source code, specifies \texttt{Requirements} for creating large and worst-case inputs, sets \texttt{Output Constraints} for both type and performance, and enforces a strict \texttt{Output Format}. A key component is the \texttt{Efficiency Example} section, which provides few-shot examples (two in our work) of the top-performing synthesizers and the execution profiles of the sampled test inputs. This feedback-driven approach is crucial for progressively synthesizing diverse and challenging test inputs to expose latent efficiency bottlenecks in LLM-translated code.

\subsection{Testscript Template}

Figures~\ref{figure:testscript_cpp}, Figure~\ref{figure:testscript_java}, and Figure~\ref{figure:testscript_python} demonstrate the designed testscript templates for C++, Java, and Python, which are engineered for automatic evaluation of code translations.

These testscripts are structured with specific placeholders to programmatically embed external code snippets during evaluation. Specifically, the \texttt{TO\_FILL\_FUNC} marker is replaced with the LLM-generated translation; the \texttt{TO\_FILL\_GOLD} marker is replaced with the ground-truth implementation from the benchmark; and the \texttt{TEST\_SUITE\_FILL} section is injected with the full suite of tests. We implemented a helper function, \texttt{areEquivalent}, to perform comparison by serializing the outputs to validate functional equivalence. During evaluation, the testscript executes and compares each test on both the LLM-generated and ground-truth translations. Finally, the script outputs a parsable summary of execution results for further analysis.

\begin{figure*}[t]
    \centering
  \begin{minipage}{1.0\textwidth}
    \begin{pinkbox}{Prompt for Stress Test Input Generation}
\begin{minted}{markdown}
### Task
You are tasked with generating a challenging test input for the following `SRC_LANG` function:
```SRC_LANG
SRC_CODE_FILL
```

### Requirements
Please carefully analyze the logic and complexity of the above source code. For each parameter in the function signature (`PARAM`):
1. Implement a Python function `generate_param_X()` for each `X`.
2. Each `generate_param_X()` must return one valid value.
3. The joint input should expose computational limits, trigger worst-case behaviour, and fully stress the function logic.

### Output Constraint
Ensure the generated test input runs within 10 seconds and 4 GB peak memory. All values must satisfy the type constraints in `SRC_LANG`.
1. Each input matches the expected type (e.g. `int`, `float`, `double`, `str`, `List[int]`, `Map<String,Integer>`,...).
2. For composite types (e.g. `vector<vector<T>>`, `pair<T1,T2>`), ensure all sub-elements conform to their declared types.

### Output Format
Implement each `generate_param_X()` and print them in order with a separator `|SEP|`. The implementation should be runnable and enclosed in a ```python...``` code block, e.g.:

```Python
import ...
def generate_param_X():
    ...
def generate_param_Y():
    ...

print(generate_param_X())
print("|SEP|")  # separator
print(generate_param_Y())
```

### Efficiency Example
Below are two validated generators and their execution profiles from their sampled values. Study their patterns and create more stressful test generators by mutating size, distribution, ordering, and edge-case choices while keeping the same output format.

- Example of Test Input Generator
```Python
GENERATOR_FILL
```

- Sampled Test Input Value
```
VALUE_FILL
```
- Efficiency Profile on SRC_LANG Code
    [Elapsed Time] TIME_FILL Seconds
    [Peak  Memory] MEM_FILL  MB
  
...(One More Example)
\end{minted}
    \end{pinkbox}
  \end{minipage}
    \caption{The prompt template used to elicit stress test inputs.}
    \label{figure:prompt_stress_test_generation}
\end{figure*}

\begin{figure*}[th]
  \centering
    \begin{minipage}[t]{1.0\textwidth}
        \begin{bluebox}[title=\textbf{C++ Testscript Template}]
\begin{minted}{cpp}
#include <iostream>
#include <cstdlib>
#include <string>
#include <vector>
#include <fstream>
#include <iomanip>
#include <bits/stdc++.h>
//TO_FILL_IMPORT

using namespace std;
//TO_FILL_FUNC
// e.g., int addOne ( int x ) {...}

//TO_FILL_GOLD
// e.g., int f_gold ( int x ) {...}

class Test {
    // Additional helper methods here ...
    
    public:
      template <typename T1, typename T2> static bool AreEquivalent(T1 o1, T2 o2) {
        return serialize_obj_(o1) == serialize_obj_(o2);
      }
    public:
      static void Start() {
        int total_num = 0;
        int pass_num = 0;
        //TEST_SUITE_FILL
        
        std::cout << "|OUTPUT| total " << total_num << " | passed " << pass_num << std::endl;
      }
};

int main() {
  Test::Start();
  return 0;
}
\end{minted}
        \end{bluebox}
    \end{minipage}
  \caption{The C++ testscript template used in our work.}
  \label{figure:testscript_cpp}
\end{figure*}

\begin{figure*}[th]
  \centering
        \begin{minipage}[t]{\textwidth}
            \begin{bluebox}[title=\textbf{Java Testscript Template}]
\begin{minted}{java}
import java.util.*;
import java.util.stream.*;
import java.lang.*;
import java.io.*;
import java.lang.reflect.Array;
//TO_FILL_IMPORT

class MAIN {
    // Additional helper methods here ...

  public static boolean areEquivalent(Object o1, Object o2) {
    return serializeObj(o1).equals(serializeObj(o2));
  }

//TO_FILL_FUNC
// e.g. public static int addOne(int x) { ... }

//TO_FILL_GOLD
// e.g. static int f_gold ( int x ) { ... }

  public static void start() {
    int total_num = 0;
    int pass_num = 0;
    //TEST_SUITE_FILL

    System.out.println("|OUTPUT| total " + total_num +  " | passed " + pass_num );
  }
  public static void main(String[] args) {
    start();
  }
}
\end{minted}
            \end{bluebox}
        \end{minipage}
  \caption{The Java testscript template used in our work.}
  \label{figure:testscript_java}
\end{figure*}
\begin{figure*}[th]
  \centering
        \begin{minipage}[t]{\textwidth}
            \begin{bluebox}[title=\textbf{Python Testscript Template}]
\begin{minted}{python}
from collections import *
from typing import *
#TO_FILL_IMPORT

# Additional helper methods here ...

def are_equivalent(o1, o2) -> bool:
    return serialize_obj(o1) == serialize_obj(o2)

#TO_FILL_FUNC
# e.g. def addOne(x): ...

#TO_FILL_GOLD
# e.g. def f_gold ( x ): ...

def start():
    total_num = 0
    pass_num = 0
    #TEST_SUITE_FILL
    
    print(f"|OUTPUT| total {total_num} | passed {pass_num}")

if __name__ == "__main__":
    start()
\end{minted}
            \end{bluebox}
        \end{minipage}
  \caption{The Python testscript template used in our work.}
  \label{figure:testscript_python}
\end{figure*}

\end{document}